\documentclass[sigconf,natbib=true]{acmart}
\usepackage{booktabs} %
\usepackage{hyperref}
\usepackage{url}
\usepackage{multirow}
\usepackage{cleveref}
\usepackage{float}
\usepackage{balance}
\usepackage{microtype}
\usepackage{comment}
\usepackage{color}
\usepackage{graphics,graphicx}
\usepackage{epstopdf}
\usepackage{colortbl}
\usepackage{diagbox}
\usepackage{tabularx}   %
\usepackage[justification=centering]{caption}
\usepackage[inline]{enumitem}
\usepackage{array}

\usepackage{todonotes}
\presetkeys%
    {todonotes}%
    {inline,backgroundcolor=yellow}{}
\usepackage{soul}

\newcommand{\post}[0]{\noindent}

\usepackage{tikz}
\usetikzlibrary{arrows}
\usetikzlibrary{decorations.text}

\usepackage{adjustbox}

\newcommand{\anoncite}[1]{\textcolor{black}{[anon.]}}

\usepackage{abbrevs}
\newabbrev\EMCB{\textit{Ethically Minded Consumer Behavior} (EMCB)}[EMCB]
\newabbrev\SRC{\textit{Socially Responsible Consumers} (SRC)}[SRC]
\newabbrev\IR{\textit{Information Retrieval} (IR)}[IR]
\newabbrev\IIR{\textit{Interactive Information Retrieval} (IIR)}[IIR]

\usepackage[utf8]{inputenc}
\DeclareUnicodeCharacter{2212}{-}

\newcolumntype{D}[1]{>{\centering\arraybackslash}p{#1}}
\newcolumntype{L}[1]{>{\arraybackslash}p{#1}}
\newcolumntype{Y}{>{\centering\arraybackslash}X}

\AtBeginDocument{%
  \providecommand\BibTeX{{%
    \normalfont B\kern-0.5em{\scshape i\kern-0.25em b}\kern-0.8em\TeX}}}

\makeatletter

\renewcommand\maybe@space@{%
  \maybe@ictrue %
  \expandafter   \@tfor
    \expandafter \reserved@a
    \expandafter :%
    \expandafter =%
                 \nospacelist
                 \do \t@st@ic
  \ifmaybe@ic %
    \space
  \fi
}

\makeatother

\renewcommand{\todo}[1]{}

\begin{document}

\copyrightyear{2025}
\acmYear{2025}
\setcopyright{cc}
\setcctype{by}
\acmConference[SIGIR '25]{Proceedings of the 48th International ACM SIGIR Conference on Research and Development in Information Retrieval}{July 13--18, 2025}{Padua, Italy}
\acmBooktitle{Proceedings of the 48th International ACM SIGIR Conference on Research and Development in Information Retrieval (SIGIR '25), July 13--18, 2025, Padua, Italy}\acmDOI{10.1145/3726302.3730347}
\acmISBN{979-8-4007-1592-1/2025/07}

\title{From Query to Conscience: The Importance of Information Retrieval in Empowering Socially Responsible Consumerism}

\author{Frans van der Sluis}
\orcid{0000-0002-3638-0784}
\email{f.vandersluis@acm.org}
\affiliation{%
  \institution{University of Copenhagen}
  \city{Copenhagen}
  \country{Denmark}
}

\author{Leif Azzopardi}
\orcid{0000-0002-6900-0557}
\email{leifos@acm.org}
\affiliation{%
  \institution{University of Strathclyde}
  \city{Glasgow}
  \country{United Kingdom}
  }

\author{Florian Meier} %
\orcid{0000-0001-9408-0686}
\email{fmeier@hum.aau.dk}
\affiliation{%
  \institution{Aalborg University}
  \city{Copenhagen}
  \country{Denmark}
  }

\begin{abstract}
    Millions of consumers search for products online each day, aiming to find items that meet their needs at an acceptable price. While price and quality are major factors in purchasing decisions, ethical considerations increasingly influence consumer behavior—giving rise to the socially responsible consumer.
Insights from a recent survey of over 600 consumers reveal that many barriers to ethical shopping stem from information-seeking challenges, often leading to decisions made under uncertainty. These challenges contribute to the intention–behaviour gap, where consumers’ desire to make ethical choices is undermined by limited or inaccessible information and inefficacy of search systems in supporting responsible decision-making. 
In this perspectives paper, we argue that the field of Information Retrieval (IR) has a critical role to play by empowering consumers to make more informed and more responsible choices. We present three interrelated perspectives: (1) reframing responsible consumption as an information extraction problem aimed at reducing information asymmetries; (2) redefining product search as a complex task requiring interfaces that lower the cost and burden of responsible search; and (3) reimagining search as a process of knowledge calibration that helps consumers bridge gaps in awareness when making purchasing decisions.
Taken together, these perspectives outline a path from query to conscience --- one where IR systems help transform everyday product searches into opportunities for more ethical and informed choices. We advocate for the development of new and novel IR systems and interfaces that address the intricacies of socially responsible consumerism, and call on the IR community to build technologies that make ethical decisions more informed, convenient, and aligned with economic realities.
\end{abstract}



\maketitle

\section{Introduction}
Within \IR, online shopping is one of the main commercially focused search tasks~\cite{Broder2002ASearch,Jansen2008DeterminingQueries}. 
Search enables consumers to efficiently explore the market by revealing what products are available, where they can be purchased, and at what price~\cite{Ghose2014ExaminingRevenue}.
Consequently, transactional search intent has played a significant role in driving the success of major web search engines and e-commerce platforms~\cite{Kallumadi2023ECom23:ECommerce}.
Within this multi-sided search market, (re)sellers, reviewers, advertisers, and producers are all incentivized
to show (\textit{more}) ads to sell (\textit{more}) products to (\textit{more}) people to make (\textit{more}) money~\cite{Tsagkias2021ChallengesRecommendations}. 
This potential for profit has attracted much attention from the IR community, investigating challenges associated with e-commerce and product search.
In~\cite{Tsagkias2021ChallengesRecommendations}, Tsagkias et al. listed the major IR related challenges in eCommerce as being centred on and around the core business-sided problem of selling -- where the focus of research is on how companies can more efficiently and effectively deliver advertisements, obtain greater engagement, and sell more products (see the ACM SIGIR eCom Workshop Series (2017-2023)~\cite{Kallumadi2023ECom23:ECommerce}). 
Conversely, little emphasis has been placed upon the user-sided problems that responsible consumers face when searching for products~\cite{azzopardi2024src}. 

Recently there has been a number of efforts such as the \textit{IR4Good} initiative\footnote{Information Retrieval for Good track, at the 46th European Conference on Information Retrieval (ECIR) in Glasgow}, or the RecSoGood initiative\footnote{
Recommender Systems for Sustainability and Social Good, workshop held in conjunction with the 19th ACM Recommender Systems (RecSys) Conference in Bari} that have been advocating for the community to consider more than just commercial interests, but the societal and environmental impacts~\cite{Felfernig2023,jannach2024rs4good} and to do ``\textit{research that matters}''~\cite{jannach2024rs4good}. For example, there has been a major push towards developing greener and more sustainable IR systems~\cite{azzopardi2009greener_search, Scells2022,Spillo2023,Vente2024}, and to create fairer and less biased  systems~\cite{azzopardi2008retrievability,beiga2018equity,wilkie2014bias,white2013biases}. In this perspectives paper, we aim to bring the community's attention to the user-sided search problems associated with online shopping. We believe that IR can and should be used to empower consumers and help them find (and purchase) products that also align with their beliefs and values~\cite{Wiederhold2018EthicalIndustry,azzopardi2024src}. 
We argue that IR can play a leading role in driving societal change by focusing on developing systems and applications which support users in being socially aware and responsible consumers.
Such consumers want to spend their money on products that align with their ethical views regarding the environment, human and animal rights, community involvement, social justice and governance~\cite{Jones2022TheDifference}. By doing so, socially responsible consumers can make a societal impact by using their purchasing power to reward companies that align with their values and penalize companies that do not.

However, while numerous movements have been advocating for different causes such as fair trade, sustainability, environmentally friendly, animal rights, diversity and inclusion, etc., more often than not, the increase in awareness does not lead to or create changes in behavior. This is often attributed to the intention-behavior gap -- where many consumers want to purchase products that align with their values, but due to the numerous challenges involved they fail to do so.
This is because, even if consumers want to find products aligned with their values, the information seeking task is difficult and complex requiring a lot of dedication, time and effort~\cite{azzopardi2024src,vandersluis2025chiir}. 
The problem is exacerbated by search engines and e-commerce platforms that are incentivized to advertise and sell ads and products which garner the greatest profit~\cite{Gleason2024,Mager2012} and producers and manufacturers that are incentivized to virtue signal, greenwash and avoid being transparent about their processes (and products) to maximise sales~\cite{shchory2020information,yang2020greenwashing,spaniol2024greenwashing}.
More often than not, what is recommended and returned through search systems and platforms is not necessarily the products that are likely to align with the user's values (by default) --  moreover platforms will often lack the options for consumers to even express or filter on aspects they care about~\cite{azzopardi2024src}. 
Here, we aim to:
\begin{enumerate}
\item describe the challenges that consumers face when trying to shop ethically and socially responsibly;
\item argue that the field of \IR can and should play a greater role in addressing these challenges, and;
\item demonstrate how these perspectives can be situated within and contribute to the broader IR research agenda on user-centered, socially responsible systems that empower users in making more informed purchasing decisions.
\end{enumerate}

The perspectives we share in this paper differ from previous discussion papers~\cite{Felfernig2023,zhou2024,jannach2024rs4good} as it is a focused, empirical and evidence-based account on how socially responsible consumers struggle with fulfilling their needs and aligning their values when shopping online. We conceptualize challenges, synthesize ongoing research on how others address these challenges and summarise our perspectives on how research in these directions should be linked to close the \textit{intention-behavior-gap} in socially responsible consumerism. 

\section{Background} \label{sec_background}
This section presents and positions our work within the relevant background on the consumer (search) journey, the challenges faced by socially responsible consumers, and the intention-behavior gap.

\subsection{Consumer Purchasing Behaviour}\label{sec_back_consumer_behaviour}
\begin{figure}
\includegraphics[width=4cm]{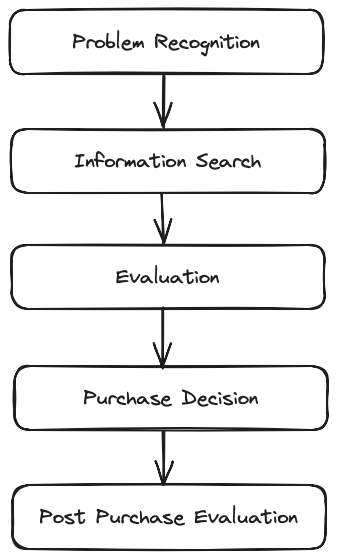}
\caption{ \citet{Engel1990CustomerBehavior}'s consumer purchasing model.}\label{fig:buying_process}
\vspace{-4mm}
\end{figure}

Most purchasing decisions involve searching for information about available products, followed by evaluating this information to make a purchase decision. This has lead to the development of a number of similar models ascribing consumer purchasing behaviour~\citep{Bloch1986ConsumerFramework,Engel1990CustomerBehavior,RowleyProductPropositions,Punj1983AnMaking,Ke2016SearchProducts,Schmidt1996ASearch}. 
According to \citep{Engel1990CustomerBehavior}, consumers go through the following stages (as shown in Figure~\ref{fig:buying_process}):

\begin{itemize}
\item \textbf{Need Recognition}: The ``awareness'' stage of the buying process arises due to some stimulus (internal or external) which results in a consumer becoming aware that they have a need or desire for a product. 
\item \textbf{Information Search}: Once a consumer recognizes that they have a need, they will often undertake a search for potential options and alternatives to help.  This stage may involve consumers simply drawing upon their internal, previous knowledge (of the product space), as well as drawing upon external information and researching their options.
\item \textbf{Evaluation of Alternatives}: During the ``consideration'' stage a consumer compares among options to make the best possible choice (or at least aims to given their constraints and concerns) based on the information that they have obtained.
\item \textbf{Purchasing Decision}: During the ``conversion'' stage, consumers turn behavior in action, and go to buy the selected option. However, consumers may still abandon their purchase and go back to searching and researching for various reasons (other alternatives are recommended, new information comes to light, etc.).
\item \textbf{Post-Purchase Evaluation}: After making a purchase, consumers will often reflect on whether their decision was worth it, whether they may consider recommending the product/brand to others, writing a review, and whether they would buy the product again or from the same brand again.
\end{itemize}

Given that searching, evaluating and comparing products and the associated information can be quite costly (in terms of time and effort), consumers frequently make decisions based on incomplete and imperfect information~\citep{WardInternetBranding,Ke2016SearchProducts}. 
And, while the availability of information about products has greatly increased, this tends to exacerbate rather than alleviate the problem for several reasons:

\begin{enumerate}
  \item Searching involves non-trivial navigation through a large variety of complex websites, which can be particularly frustrating and cognitively taxing \citep{Brynjolfsson1999FrictionlessRetailers}.
  \item Information asymmetries exist between parties, where the seller or producer don't disclose pertinent information about their products \citep{Akerlof1970TheMechanism,Jacoby1974BrandLoad}.
  \item With many alternative products to choose from, and each with varying attributes, consumers often experience choice and decision overload \citep{Keller1987EffectsEffectiveness}.
 \end{enumerate}
So rather than helping, more information often decreases consumers' decision-making effectiveness \citep{Keller1987EffectsEffectiveness,Schleenbecker2015InformationCoffee}. Though, branding \citep{WardInternetBranding} and labeling \citep{Young2010SustainableProducts,Annunziata2011ConsumersProducts} (e.g. Fair Trade, Energy Efficiency, etc.) have been proposed as solutions to help reduce information asymmetry and facilitate more efficient decision making as consumers tend to prefer general information over specific information~\cite{Schleenbecker2015InformationCoffee}. When confronted with ethical information during the search process, consumers often become more concerned and perform more extensive searching~\cite{Zander2012InformationFood}. However, the majority of consumers tend to use simplifying and selective search strategies when looking for ethical information~\cite{Zander2012InformationFood}.

\subsection{Socially Responsible Consumers}\label{bg:src}
A ``\textit{socially responsible consumer}'' is an individual who makes purchasing decisions with a focus on the broader social and environmental impact of their choices, in addition to their personal needs and preferences~\cite{Ellen1991TheBehaviors,CarringtonLostGap,Jones2022TheDifference,Davies2016ConsumerConsumption, Sheth2011MindfulSustainability}\footnote{Socially responsible consumers are also referred to as a mindful consumers or ethical consumers.}. 
According to Carrigan et al.~\cite{carrigan2004better_shopping}, \textit{responsible consumerism} broadly stands for ``\textit{the conscious and deliberate choice to make certain consumption choices due to personal and moral beliefs}'' and thus responsible consumers prioritize products and services that align with their personal values and beliefs~\cite{CarringtonLostGap,Jones2022TheDifference}. 
These may include beliefs and values about the environment, human and animal rights, community involvement, social justice, governance, etc.~\cite{Jones2022TheDifference,Shaw2006FashionChoice}.

Social responsibility, however, is perceived differently by each consumer, meaning that an aspect important to one person might not be important to another. Hasanzade et al.~\cite{Hasanzade2018SelectingShopping} surveyed 249 consumers in Germany and found that most participants were ethically minded consumers (54\%), while the other participants were concerned with price (12\%) and price-quality (34\%). Of those ethically minded, most were concerned about animal rights, followed by labour/human rights, and then environment protection. Casais et al.~\cite{Casais2022ThePriorities} surveyed 364 consumers in Portugal, of which most considered themselves as socially responsible consumers, reported that they were concerns about labour/human rights (31\% ), environmental issues (23\%), animal rights/welfare (17\%), and all three (27\%). The variety of ethical concerns translates into highly varied purchasing decisions.  For example, a consumer who consciously avoids products made in regions associated with slave/forced labour may be indifferent to the ecological or sustainability implications of their choice. Therefore, not all consumers will attribute the same importance to the different issues/aspects associated with being ethical and socially responsible.

Nonetheless, socially responsible consumers aim to use their purchasing power as a means to promote positive change. 
They believe that by supporting companies and products that align with their values and sustainability goals, they can influence businesses to be more socially and environmentally responsible~\cite{Liu2018TheIntention}. 
This consumer mindset has grown in prominence with increasing awareness of global environmental issues, labor rights, and corporate responsibility, leading to the growth of various certifications and labels to help consumers identify products that meet these criteria~\cite{Jones2022TheDifference}. 
Socially responsible consumers tend to establish an identity rooted in their ethical purchasing decisions, occasionally making personal sacrifices~\cite{Papaoikonomou2018LookingApproach}. 
And, they often tend to communicate their role as advocates for a more sustainable society to others~\cite{Casais2022ThePriorities}.

\subsection{Intention-Behaviour Gap}
The \textit{Theory of Planned Behaviour} seeks to provide an explanation of behaviour by considering how the individual's attitudes and norms influences their intentions and subsequently their behaviour in a causal sequence~\cite{Ajzen1985FromBehavior}.
People's intentions to purchase socially responsible products, however, do not always align with their behaviors, creating a \textit{Intention-Behavior Gap}~\cite{Ajzen1985FromBehavior}. 
For example, of 81 self-declared green consumers, 30\% reported that they were very concerned about environmental issues but they struggled to translate this into their purchasing decisions~\cite{Young2010SustainableProducts}. 
Within the literature on consumer purchasing, this misalignment has been of great interest, especially regarding ethical concerns~\cite{Moon2004ConsumerHypotheses,Ozcaglar-Toulouse2006InFrance,RowleyProductPropositions}. 

Uusitalo et al.~\cite{Uusitalo2004EthicalFinland} conducted a study of Finnish consumers (n=713) and found that while the majority of the participants regarded ethics as important, it did not necessarily lead to ethical choices regarding purchases. They found participants were uncertain about which products and which companies were ethical and acting socially responsibly. The major obstacles to being a socially responsible consumer were difficulties in obtaining information, problems in product availability (i.e., lack of ethical/responsible options to select from), and higher prices of ethical products. In a study on sustainable fashion with German participants ($n=13$), they found that the following barriers impeded ethical purchasing decisions: higher perceived prices, lack of availability and fewer ethical items to select from (restricting their ``image''), lack of knowledge and education about ethical aspects, lack of information about the different aspects, moreover the lack of transparency and credibility of (labeling) information, 
coupled with current consumption habits and inertia to change (and if they change whether it would make a difference anyway)~\cite{Wiederhold2018EthicalIndustry}.

In a nine-month in-depth study of people's purchasing behaviors (n=13), Carrington et al~\cite{CarringtonLostGap} found that four interrelated factors affecting the ethical intention-behavior gap: 
\begin{enumerate*}
    \item prioritization of ethical concerns, 
    \item formation of plans/habits,
    \item willingness to commit and sacrifice, and, 
    \item modes of shopping behavior. 
\end{enumerate*}
They found that obstacles such as alternative personal values, persistent habits, lack of planning, unwillingness to sacrifice, missing information, and situational distractions often prevent these factors from translating into ethical action.

Djafarova et al.~\cite{Djafarova2022ExploringBehaviour} interviewed a cohort of participants from Generation Z based in the UK ($n=18$) and found that they had strong awareness and desire toward ethical and environmental concerns. However, they felt limited by their finances when considering high-value items and instead tended to exercise more responsibility by recycling, dieting choices, and reducing consumption.

More recently, \citet{azzopardi2024src} conducted a survey ($n=286$) investigating the role of search in the intention-behaviour gap.
They found that there was a progressive gap from how much importance participants placed on different aspects pertaining to a product, whether they considered such aspects during their purchase decision, and finally to whether they choose to search for more information about the aspects (or not).
Participants who did search for secondary information reported challenges including low awareness, limited accessibility, difficulties in finding and understanding relevant details, lack of trust, and the complexity of comparing alternatives.

In a follow up task based study ($n=308$), they found that searching for ethical information can increase the perceived importance of ethical considerations in purchasing decisions, however, this effect was not primarily driven by pre-existing ethical intentions or the act of searching. Instead, meaningful behavior change occurred when consumers recognized gaps in their knowledge and were able to make sense of the ethical information they encountered. These findings suggest that responsible consumption is to certain extent an information problem, where increasing awareness and comprehension plays a pivotal role in aligning consumer choices with ethical values.

\section{Survey on Search-Related Obstacles to Responsible Consumerism}
To investigate the challenges consumers face when trying to make purchasing decisions we expanded upon past work \citep{azzopardi2024src} surveying a larger and broader range of participants.
The survey instrument was designed to evaluate the extent to which search supported or impeded participants' ability to make informed purchasing decisions. 
The survey asked participants to recall a recent purchase valued at least $100\$$ and asked them about their reasons (not) to search\footnote{\scriptsize{Ethics approval for the study was granted  by the University of Strathclyde's Department of Computer and Information Sciences Ethics Committee (Application No. 2294).}.}. 
We also asked participants about their reasons for why they did not search for information regarding the range of socially responsible aspects mentioned above (e.g., labor and employment rights, fair trade, environmental friendly, sustainability, social justice, ideology, etc.).
For full details of the survey, see \citep{azzopardi2024src}.

In this work, we report our findings based on $601$ participants to provide a more representative sample of consumers (compared with 286 in \citep{azzopardi2024src}). 
The extended sample had a similar distribution for age (mean: 33.55, std: 11.51), gender (370 male, 220 female, 11 unknown) and geography (top-5: United Kingdom 198, Poland 83, South Africa 57, United States 56, and Portugal 33) as in~\citep{azzopardi2024src}.
The greater sample size enables us to examine different consumer cohorts 
based their \EMCB questionnaire scores.
\begin{figure}[tbh]
    \centerline{ \scalebox{0.9}{\input{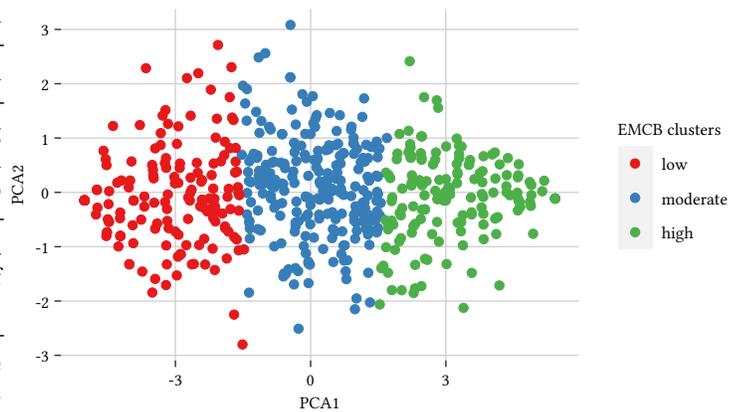}}}
\vspace{-.5cm}
\caption{2D Scatter plot of participants with K-Means Clustering based on their EMCB answers. K-Means cluster analysis was performed with 3 clusters and after scaling answers to standardized scores (mean and std.). Plot axes are derived from a Principal Component Analysis (PCA). The figure indicates a clear distinction between clusters on the major principal component (PCA1, $R^2=64.44\%$), whilst variability remained on the second component (PCA2, $R^2=7.44\%$).
}\label{fig:clusters}
\end{figure}

\subsection{From Ethically Minded to Ethically Agnostic}\label{sec_emcb}
The \EMCB is a validated questionnaire that measures participants' intentions towards ethical consumption~\cite{SUDBURYRILEY20162697}, including ecological and social-ethical concerns and willingness to recycle and pay more.
Given the survey results, we applied factor analysis to identify different types of consumers. The elbow method was used to determine the appropriate number of clusters. Figure \ref{fig:clusters} depicts the three clusters found.
Of our 601 participants, $29.09\%$ were low on the \EMCB scale (with a mean of $1.74)$  suggesting that they were the least ethically minded and/or ethically agnostic showing little intention or regard for purchasing ethical and socially responsible products.
$43.72\%$ were moderately ethically minded (with a mean of $2.98$), expressing a mixture of ethically mindedness towards certain aspects of being socially responsible (but not all). For example, they may have expressed concerns over environmental aspects but not employment rights, or vice versa.
The remaining $27.19\%$ were considered to be highly ethically consumers (with a mean of $4.13$), showing much concern and demonstrated behaviours towards shopping ethically and responsibly. For example, such participants indicated that they switched to more environmentally friendly products, used recycled goods/containers, avoided products that caused harm, avoided socially irresponsible companies, etc.
This analysis suggests considerable differences between groups of participants in their intentions towards responsible consumption.
The outcome falls in line with previous works that found similar distributions of ethically minded consumers~\cite{Casais2022ThePriorities,Hasanzade2018SelectingShopping}.

\begin{figure*}
    \centerline{\input{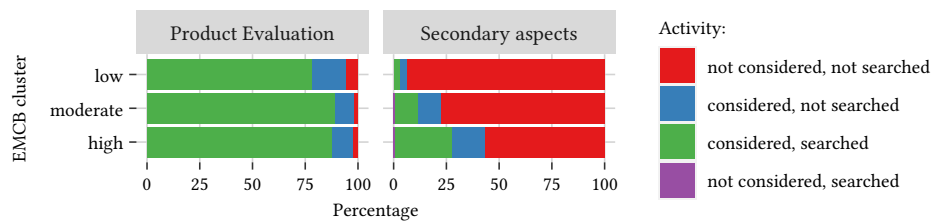}}
\vspace{-0.5cm}
\caption{%
Influence of EMCB cluster on participants' intention-search gap. The gap is expressed in terms of
participants' activities during their purchase decision; whether they considered and/or searched for an aspect.
Aspects are grouped per theme, where the themes of 'Environmental and Social Responsibility', 'Ethical Procurement', and 'Inclusion and Ideology' are combined under 'Secondary aspects'.
Chi-square tests showed a small, significant influence of EMCB cluster for Product Evaluation aspects ($\chi^2 = 18.03$, $\textit{df} = 6$, $p < .01$).
For secondary aspects, a large, significant influence of EMCB was confirmed ($\chi^2 = 218.87$, $\textit{df} = 6$, $p <.001$).
}
\label{fig:searches}
\end{figure*}

Figure ~\ref{fig:searches} displays the extent to which intentions extended into considerations and search activity.
For product-related criteria, which are primary to the purchase decision, the majority of participants indicated that they considered and also searched for these criteria. Chi-square analysis detailed in Figure~\ref{fig:searches} showed that their chances of acting upon product criteria was slightly affected by their EMCB cluster, suggesting a general tendency towards more considered decision making.
For aspects related to social-environmental responsibility and ethics, a minority of participants considered or searched.
Chi-square analysis shows that the probability of considering these aspects depended significantly on \EMCB cluster, ranging from $6.04\%$ for low \EMCB, $21.52\%$ for moderate \EMCB, to $42.42\%$ for high \EMCB.
When considered, the odds of searching furthermore increased from $39.39\%$, $51.14\%$, to $62.95\%$ with \EMCB scores.
These results indicate that intentions partly explain subsequent considerations and search activity.
Nevertheless, even for the high \EMCB cluster,
a considerable gap remained with $73.30\%$ either not considering or not searching for these aspects.

\subsection{Reasons Not to Search}
\begin{figure*}
    \centerline{\begin{tikzpicture}[x=1pt,y=1pt]
\definecolor{fillColor}{RGB}{255,255,255}
\path[use as bounding box,fill=fillColor,fill opacity=0.00] (0,0) rectangle (433.62,144.54);
\begin{scope}
\path[clip] (  0.00,  0.00) rectangle (433.62,144.54);

\path[] (  0.00, -0.00) rectangle (433.62,144.54);
\end{scope}
\begin{scope}
\path[clip] (110.93, 27.33) rectangle (354.98,139.04);

\path[] (110.93, 27.33) rectangle (354.98,139.04);
\definecolor{drawColor}{RGB}{211,211,211}

\path[draw=drawColor,line width= 0.6pt,line join=round] (110.93, 35.50) --
	(354.98, 35.50);

\path[draw=drawColor,line width= 0.6pt,line join=round] (110.93, 49.13) --
	(354.98, 49.13);

\path[draw=drawColor,line width= 0.6pt,line join=round] (110.93, 62.75) --
	(354.98, 62.75);

\path[draw=drawColor,line width= 0.6pt,line join=round] (110.93, 76.37) --
	(354.98, 76.37);

\path[draw=drawColor,line width= 0.6pt,line join=round] (110.93, 90.00) --
	(354.98, 90.00);

\path[draw=drawColor,line width= 0.6pt,line join=round] (110.93,103.62) --
	(354.98,103.62);

\path[draw=drawColor,line width= 0.6pt,line join=round] (110.93,117.24) --
	(354.98,117.24);

\path[draw=drawColor,line width= 0.6pt,line join=round] (110.93,130.87) --
	(354.98,130.87);

\path[draw=drawColor,line width= 0.6pt,line join=round] (122.03, 27.33) --
	(122.03,139.04);

\path[draw=drawColor,line width= 0.6pt,line join=round] (177.59, 27.33) --
	(177.59,139.04);

\path[draw=drawColor,line width= 0.6pt,line join=round] (233.16, 27.33) --
	(233.16,139.04);

\path[draw=drawColor,line width= 0.6pt,line join=round] (288.73, 27.33) --
	(288.73,139.04);

\path[draw=drawColor,line width= 0.6pt,line join=round] (344.30, 27.33) --
	(344.30,139.04);
\definecolor{fillColor}{RGB}{55,126,184}

\path[fill=fillColor] (122.03,128.82) rectangle (299.11,132.91);

\path[fill=fillColor] (122.03,115.20) rectangle (276.04,119.29);
\definecolor{fillColor}{RGB}{228,26,28}

\path[fill=fillColor] (122.03, 97.49) rectangle (343.89,101.58);

\path[fill=fillColor] (122.03,124.74) rectangle (320.48,128.82);

\path[fill=fillColor] (122.03,111.11) rectangle (319.47,115.20);
\definecolor{fillColor}{RGB}{77,175,74}

\path[fill=fillColor] (122.03,132.91) rectangle (276.73,137.00);
\definecolor{fillColor}{RGB}{55,126,184}

\path[fill=fillColor] (122.03, 87.95) rectangle (212.26, 92.04);

\path[fill=fillColor] (122.03, 74.33) rectangle (198.69, 78.42);

\path[fill=fillColor] (122.03,101.58) rectangle (191.23,105.66);

\path[fill=fillColor] (122.03, 60.71) rectangle (188.52, 64.79);
\definecolor{fillColor}{RGB}{228,26,28}

\path[fill=fillColor] (122.03, 83.87) rectangle (214.64, 87.95);
\definecolor{fillColor}{RGB}{55,126,184}

\path[fill=fillColor] (122.03, 47.08) rectangle (180.38, 51.17);

\path[fill=fillColor] (122.03, 33.46) rectangle (176.98, 37.55);
\definecolor{fillColor}{RGB}{77,175,74}

\path[fill=fillColor] (122.03, 51.17) rectangle (190.43, 55.26);

\path[fill=fillColor] (122.03, 64.79) rectangle (189.38, 68.88);

\path[fill=fillColor] (122.03,119.29) rectangle (189.38,123.37);

\path[fill=fillColor] (122.03, 92.04) rectangle (185.17, 96.13);
\definecolor{fillColor}{RGB}{228,26,28}

\path[fill=fillColor] (122.03, 70.24) rectangle (181.05, 74.33);
\definecolor{fillColor}{RGB}{77,175,74}

\path[fill=fillColor] (122.03, 78.42) rectangle (176.75, 82.50);
\definecolor{fillColor}{RGB}{228,26,28}

\path[fill=fillColor] (122.03, 29.37) rectangle (166.81, 33.46);

\path[fill=fillColor] (122.03, 56.62) rectangle (155.61, 60.71);
\definecolor{fillColor}{RGB}{77,175,74}

\path[fill=fillColor] (122.03, 37.55) rectangle (155.70, 41.63);

\path[fill=fillColor] (122.03,105.66) rectangle (154.65,109.75);
\definecolor{fillColor}{RGB}{228,26,28}

\path[fill=fillColor] (122.03, 43.00) rectangle (146.45, 47.08);
\end{scope}
\begin{scope}
\path[clip] (  0.00,  0.00) rectangle (433.62,144.54);
\definecolor{drawColor}{RGB}{0,0,0}

\node[text=drawColor,anchor=base east,inner sep=0pt, outer sep=0pt, scale=  0.80] at (105.98, 32.75) {8. No good alternatives};

\node[text=drawColor,anchor=base east,inner sep=0pt, outer sep=0pt, scale=  0.80] at (105.98, 46.37) {7. Already knew needed info};

\node[text=drawColor,anchor=base east,inner sep=0pt, outer sep=0pt, scale=  0.80] at (105.98, 60.00) {6. Information hard to find};

\node[text=drawColor,anchor=base east,inner sep=0pt, outer sep=0pt, scale=  0.80] at (105.98, 73.62) {5. Too time consuming};

\node[text=drawColor,anchor=base east,inner sep=0pt, outer sep=0pt, scale=  0.80] at (105.98, 87.24) {4. Can't take action};

\node[text=drawColor,anchor=base east,inner sep=0pt, outer sep=0pt, scale=  0.80] at (105.98,100.86) {3. Don't care};

\node[text=drawColor,anchor=base east,inner sep=0pt, outer sep=0pt, scale=  0.80] at (105.98,114.49) {2. Wouldn't impact decision};

\node[text=drawColor,anchor=base east,inner sep=0pt, outer sep=0pt, scale=  0.80] at (105.98,128.11) {1. Not considered before};
\end{scope}
\begin{scope}
\path[clip] (  0.00,  0.00) rectangle (433.62,144.54);
\definecolor{drawColor}{gray}{0.20}

\path[draw=drawColor,line width= 0.6pt,line join=round] (108.18, 35.50) --
	(110.93, 35.50);

\path[draw=drawColor,line width= 0.6pt,line join=round] (108.18, 49.13) --
	(110.93, 49.13);

\path[draw=drawColor,line width= 0.6pt,line join=round] (108.18, 62.75) --
	(110.93, 62.75);

\path[draw=drawColor,line width= 0.6pt,line join=round] (108.18, 76.37) --
	(110.93, 76.37);

\path[draw=drawColor,line width= 0.6pt,line join=round] (108.18, 90.00) --
	(110.93, 90.00);

\path[draw=drawColor,line width= 0.6pt,line join=round] (108.18,103.62) --
	(110.93,103.62);

\path[draw=drawColor,line width= 0.6pt,line join=round] (108.18,117.24) --
	(110.93,117.24);

\path[draw=drawColor,line width= 0.6pt,line join=round] (108.18,130.87) --
	(110.93,130.87);
\end{scope}
\begin{scope}
\path[clip] (  0.00,  0.00) rectangle (433.62,144.54);
\definecolor{drawColor}{gray}{0.20}

\path[draw=drawColor,line width= 0.6pt,line join=round] (122.03, 24.58) --
	(122.03, 27.33);

\path[draw=drawColor,line width= 0.6pt,line join=round] (177.59, 24.58) --
	(177.59, 27.33);

\path[draw=drawColor,line width= 0.6pt,line join=round] (233.16, 24.58) --
	(233.16, 27.33);

\path[draw=drawColor,line width= 0.6pt,line join=round] (288.73, 24.58) --
	(288.73, 27.33);

\path[draw=drawColor,line width= 0.6pt,line join=round] (344.30, 24.58) --
	(344.30, 27.33);
\end{scope}
\begin{scope}
\path[clip] (  0.00,  0.00) rectangle (433.62,144.54);
\definecolor{drawColor}{RGB}{0,0,0}

\node[text=drawColor,anchor=base,inner sep=0pt, outer sep=0pt, scale=  0.80] at (122.03, 16.87) {0\%};

\node[text=drawColor,anchor=base,inner sep=0pt, outer sep=0pt, scale=  0.80] at (177.59, 16.87) {10\%};

\node[text=drawColor,anchor=base,inner sep=0pt, outer sep=0pt, scale=  0.80] at (233.16, 16.87) {20\%};

\node[text=drawColor,anchor=base,inner sep=0pt, outer sep=0pt, scale=  0.80] at (288.73, 16.87) {30\%};

\node[text=drawColor,anchor=base,inner sep=0pt, outer sep=0pt, scale=  0.80] at (344.30, 16.87) {40\%};
\end{scope}
\begin{scope}
\path[clip] (  0.00,  0.00) rectangle (433.62,144.54);
\definecolor{drawColor}{RGB}{0,0,0}

\node[text=drawColor,anchor=base,inner sep=0pt, outer sep=0pt, scale=  0.80] at (232.96,  7.06) {Frequency};
\end{scope}
\begin{scope}
\path[clip] (  0.00,  0.00) rectangle (433.62,144.54);
\definecolor{drawColor}{RGB}{0,0,0}

\node[text=drawColor,rotate= 90.00,anchor=base,inner sep=0pt, outer sep=0pt, scale=  0.80] at ( 11.01, 83.18) {Reasons};
\end{scope}
\begin{scope}
\path[clip] (  0.00,  0.00) rectangle (433.62,144.54);

\path[] (365.98, 50.47) rectangle (428.12,115.90);
\end{scope}
\begin{scope}
\path[clip] (  0.00,  0.00) rectangle (433.62,144.54);
\definecolor{drawColor}{RGB}{0,0,0}

\node[text=drawColor,anchor=base west,inner sep=0pt, outer sep=0pt, scale=  0.80] at (371.48,104.11) {EMCB cluster};
\end{scope}
\begin{scope}
\path[clip] (  0.00,  0.00) rectangle (433.62,144.54);
\definecolor{fillColor}{gray}{0.95}

\path[fill=fillColor] (371.48, 84.88) rectangle (385.94, 99.33);
\end{scope}
\begin{scope}
\path[clip] (  0.00,  0.00) rectangle (433.62,144.54);
\definecolor{fillColor}{RGB}{228,26,28}

\path[fill=fillColor] (372.20, 85.59) rectangle (385.23, 98.62);
\end{scope}
\begin{scope}
\path[clip] (  0.00,  0.00) rectangle (433.62,144.54);
\definecolor{fillColor}{gray}{0.95}

\path[fill=fillColor] (371.48, 70.43) rectangle (385.94, 84.88);
\end{scope}
\begin{scope}
\path[clip] (  0.00,  0.00) rectangle (433.62,144.54);
\definecolor{fillColor}{RGB}{55,126,184}

\path[fill=fillColor] (372.20, 71.14) rectangle (385.23, 84.17);
\end{scope}
\begin{scope}
\path[clip] (  0.00,  0.00) rectangle (433.62,144.54);
\definecolor{fillColor}{gray}{0.95}

\path[fill=fillColor] (371.48, 55.97) rectangle (385.94, 70.43);
\end{scope}
\begin{scope}
\path[clip] (  0.00,  0.00) rectangle (433.62,144.54);
\definecolor{fillColor}{RGB}{77,175,74}

\path[fill=fillColor] (372.20, 56.68) rectangle (385.23, 69.71);
\end{scope}
\begin{scope}
\path[clip] (  0.00,  0.00) rectangle (433.62,144.54);
\definecolor{drawColor}{RGB}{0,0,0}

\node[text=drawColor,anchor=base west,inner sep=0pt, outer sep=0pt, scale=  0.80] at (389.94, 89.35) {low};
\end{scope}
\begin{scope}
\path[clip] (  0.00,  0.00) rectangle (433.62,144.54);
\definecolor{drawColor}{RGB}{0,0,0}

\node[text=drawColor,anchor=base west,inner sep=0pt, outer sep=0pt, scale=  0.80] at (389.94, 74.90) {moderate};
\end{scope}
\begin{scope}
\path[clip] (  0.00,  0.00) rectangle (433.62,144.54);
\definecolor{drawColor}{RGB}{0,0,0}

\node[text=drawColor,anchor=base west,inner sep=0pt, outer sep=0pt, scale=  0.80] at (389.94, 60.44) {high};
\end{scope}
\end{tikzpicture}}
\caption{Participants' selections of pre-listed reasons for not searching on aspects related to \textit{Environmental and Social Responsibility}, \textit{Inclusion and Ideology}, and \textit{Ethical Procurement}. Frequencies of selections are relative to the total number of times participants were asked about an aspect. Reasons displayed may be shortened from their original descriptions.
}\label{fig:reasons}
\end{figure*}
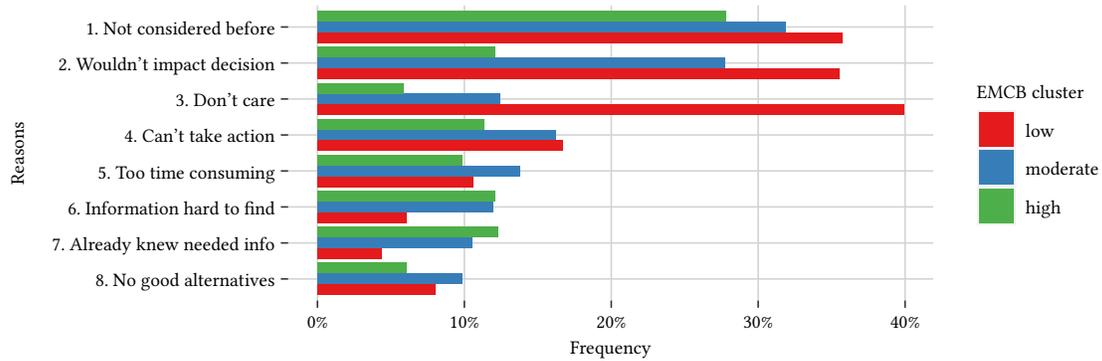

To explore why participants did not seek information on the socially responsible and ethical aspects (from here on: secondary aspects), 
we asked them to indicate their reasons for not searching. 
Counts for a pre-listed set of reasons are summarized in Figure~\ref{fig:reasons}.

The responses revealed a common theme within participants' decision-making processes: The perceived value of acquiring information related to responsible consumption aspects often fell short to the associated costs of searching. 
Participants reported multifaceted challenges, spanning from the unavailability or complexity of locating reliable information to the inadequacy of their prior knowledge to effectively search for secondary aspects, making seeking `too time consuming' and information `hard to find'.
In addition, participants questioned the relevance of secondary aspects during decision making citing their unawareness of potential issues in the supply chain, such that they had `not considered [it] before' or assumed they `already knew needed info'.
Some participants, however, cited limited motivation (`don't care', `wouldn't impact decision') rather than unawareness.
Effort-related barriers were more common among participants with moderate to high intentions, while value-related reasons were more often endorsed by those with lower intentions (Figure~\ref{fig:reasons}).

Collectively, these obstacles limited participants' capacity to engage in searches and frequently lead to \textit{a priori} neglect of secondary aspects during decision making.
Secondary aspects were often overlooked or undervalued, yet recognizing their importance often requires the very search that feels too burdensome to begin \citep{vandersluis2025chiir}.

\subsection{Information Seeking Challenges}\label{challenges}
Our survey results and literature review highlight several user-side obstacles responsible consumers face when seeking product information. 
Here, we complement the pre-listed reasons in Figure~\ref{fig:reasons} with rich survey comments that capture additional perspectives,
as well as extend previous work by re-structuring the identified challenges using the consumer buying process model (Figure~\ref{fig:buying_process}).

\paragraph{Problem Recognition}\label{cha:awareness}
As previously mentioned, the problem recognition phase is an important step in the buying process where the consumer identifies their needs. 
For the responsible consumer (or at least for a consumer considering to be responsible, ethical, etc.), additional, secondary aspects regarding what is required and relevant need to be considered.
Here, our findings indicate that consumers face challenges in:
\begin{enumerate*}
    \item realizing that secondary aspects are relevant to their purchase decisions;
    \item recognizing their relevance within the product category; and,
    \item acknowledging gaps in their own knowledge,
\end{enumerate*}
as illustrated by these quotes:
\begin{center}
\begin{tabular}{p{.9\linewidth}}
\midrule
\post ``\textit{I didn't consider this as a priority, but I could be very well convinced of the relevance of this information}''. \\
\multicolumn{1}{r}{(Country of Origin / Place of Manufacture - Participant 30)} \\
\end{tabular}
\begin{tabular}{p{1.0\linewidth}}
\post ``\textit{Not sure how it would be relevant when buying a camera}''. \\
\multicolumn{1}{r}{(Ethical Sourcing / Production - Participant 599)} \\
\end{tabular}
\begin{tabular}{p{1.0\linewidth}}
\post ``\textit{I just assumed every company is the same in this matter, so I didn't even consider it}''. \\
\multicolumn{1}{r}{(Political Stance / Ideology - Participant 30)} \\
\midrule
\end{tabular}
\end{center}

While some of these challenges may stem from consumer apathy towards responsible consumption, 
these quotes overall align with quantitative findings in Figure \ref{fig:reasons} on a lack of awareness of secondary aspects and of knowledge gaps (reasons 1 and 7). 
They indicate that an initial lack of problem recognition can prevent searching, but also show an openness of participants to re-evaluate their considerations and assumptions.

\paragraph{Information Search}\label{cha:search}
After the recognition of an (information) need, the search for information itself may pose a challenge to responsible decision making. 
Our findings indicate that consumers specifically face challenges related to:
\begin{enumerate*}
    \item the availability and existence of information regarding the different aspects; and,
    \item the accessibility of such information and the cost to obtain and find said information.
\end{enumerate*}
This was highlighted by the following quotes by participants:
\begin{center}
\begin{tabular}{p{.9\linewidth}}
\midrule
\post ``\textit{The information isn't readily available}''. \\
\multicolumn{1}{r}{(Slave/Child Labour / Employment Rights - Participant 512)} \\
\end{tabular}
\begin{tabular}{p{1.0\linewidth}}
\post ``\textit{If the information was on the product description, I would consider it, but otherwise, it was a topic that I did not check}''. \\
\multicolumn{1}{r}{(Ethical Sourcing / Production - Participant 277)} \\
\midrule
\end{tabular}
\end{center}
These challenges are echoed in reason 5 \textit{``too time consuming''} and 6 \textit{``hard to find''} in Figure \ref{fig:reasons}.
They show that, for information associated with secondary aspects to be retrieved, it must first exist, and then be accessible without being prohibitively costly to the searcher.

\paragraph{Information Evaluation}\label{cha:evaluation}
Once consumers have found a number of product alternatives and information related to those products, they need to evaluate the alternatives -- based on the information they have found. Assuming the consumer can find relevant information about the secondary aspects of concern, they face two more challenges:
\begin{enumerate*}
\item understandability of the information and how to make sense of the labeling, regulations, statistics, etc.; and,
\item reliability of the information and whether it can be trusted and used,
\end{enumerate*}
as illustrated by the following quotes: 
\begin{center}
\begin{tabular}{p{1.0\linewidth}}
\midrule
\post ``\textit{The biggest challenges when shopping and comparing products online are sifting through information overload, verifying the authenticity of reviews, and dealing with incomplete or inconsistent product details.}'' \\
\multicolumn{1}{r}{(Participant 176)} \\
\end{tabular}
\begin{tabular}{p{1.0\linewidth}}
\post ``\textit{Due to the nature of components used(rare earth metals, and metal in general) I did not conduct further research into it as this information is difficult to validate.}'' \\
\multicolumn{1}{r}{(Ethical Sourcing / Production - Participant 314)} \\
\midrule
\end{tabular}
\end{center}
These challenges were not covered by the pre-listed reasons in Figure \ref{fig:reasons}, but were nonetheless raised by participants.
They underscore the cognitive burden imposed by complex, incomplete and sometimes obscured information sources about manufacturing practices and ethical standards.

\paragraph{Purchase Decision}\label{cha:decision}

Making a purchasing decisions is complex, involving a multitude of primary aspects such as product features, quality, and price considerations. Responsible purchasing decisions introduce several additional layers of complexity, encompassing the secondary aspects associated with ethical, environmental, and social considerations. 
Here, our findings indicate that consumers face challenges in:
\begin{enumerate*}
\item dealing with overload of information and alternatives;
\item dealing with the incompleteness and uncertainty associated with the information found (and not found); and,
\item making trade-offs between alternatives on both primary and secondary aspects.
\end{enumerate*}
\begin{center}
\begin{tabular}{p{1.0\linewidth}}
\midrule
\post ``\textit{I should have looked further but became very overwhelmed with making this decision}'' \\
\multicolumn{1}{r}{(Participant 473)} \\
\end{tabular}
\begin{tabular}{p{1.0\linewidth}}
\post ``\textit{Sometimes there is too much choice which can confuses matters and takes up more time to determine the best option.}'' \\
\multicolumn{1}{r}{(Participant 115)} \\
\midrule
\end{tabular}
\end{center}
These comments highlight how the broader range of aspects and criteria considered, the challenge of making well-calibrated decisions becomes even more intricate, time-consuming, time-pressured and difficult.

\paragraph{Conclusion}
The array of challenges that users face are not experienced in isolation, but rather in junction with each other as consumers progress through the stages of the consumer search process.
The challenge begins with considering secondary aspects, recognizing their general importance, and their specific relevance to the product category.
Once considered, consumers need awareness of knowledge gaps
and translating those into search queries.
When available, information found can be unreliable, biased, or difficult to understand.
And finally, even with complete information, consumers face the challenge of weighing primary aspects (e.g., price, features, quality) against secondary aspects (e.g., environmental, social, ethical considerations), leading to difficult trade-offs~\cite{CarringtonLostGap,Flavian-Blanco2011AnalyzingEngines}. 
This complexity can deter consumers, as the perceived value of responsible consumption information often does not justify the search effort. 
Conversely, awareness of this complexity can undermine confidence in their search and decision-making abilities, potentially leading to an \textit{a priori} neglect of responsible aspects~\cite{alba2000}.
Taken together, these search challenges underscore the multifaceted and likely prohibitive nature of the seeking and decision-making tasks involved in responsible consumption.

\section{Perspectives}

In this section, we consider how the different information seeking challenges interact while providing a number of perspectives that highlight the importance of \IR along with ongoing efforts on how to address said challenges.

\subsection{An Information Extraction Problem}\label{sec:socio}
When shopping online, consumers face many information asymmetries — situations where sellers or producers withhold or fail to disclose important details about their products, leaving consumers without the full information needed to make informed decisions~\citep{Akerlof1970TheMechanism,Jacoby1974BrandLoad}.
This disparity also arises from selective disclosure and green-washing tactics employed by companies, creating an illusion of responsibility while obscuring negative aspects~\cite{deFreitasNetto2020ConceptsReview,spaniol2024greenwashing}. 
Additionally, complex global supply chains make it difficult for consumers to track a product's ethical journey. Inconsistencies in data quality and verification further obscure transparency~\cite{Wiederhold2018EthicalIndustry}, whilst
consumers may struggle to comprehend technical terms and presentations ~\cite{Wiederhold2018EthicalIndustry}.
These asymmetries can distort consumers’ perceptions of products and brands, leading to misinformed evaluations and undermining their ability to make ethically sound choices.
Such tactics not only limit access to reliable information (Section \ref{challenges}, \textit{\nameref{cha:search}}), but also overwhelm it with noise, shifting the burden of verification onto consumers — if they can find any accurate information at all (Section \ref{challenges}, \textit{\nameref{cha:evaluation}}).

The challenges related to the availability, accessibility and quality of information on secondary aspects are partly rooted in existing economic arrangements. 
Information contributing to product sales is readily available, but information scrutinizing those products is not.
Non-profit organizations, particularly NGOs, have been instrumental in addressing this imbalance. 
They provide branding and labeling schemes, such as Fair Trade and Energy Efficiency labels, aimed at improving information availability ~\citep{WardInternetBranding,Young2010SustainableProducts,Annunziata2011ConsumersProducts}.
Additionally, they provide fact-checking services that aim at improving information reliability~\citep{brandtzaeg2017}.
However, these endeavors often require extensive efforts, primarily led by specialized organizations, leading to a fragmented information landscape that consumers need to understand and navigate. 
We propose that IR can significantly contribute to these efforts by:
\begin{description}
    \item[Product information on secondary aspects] 
    Providing the technology to generate, extract and synthesize information on secondary aspects of products from various often disparate sources. 
    The research challenge lies in effectively integrating these sources at scale, while ensuring the quality required for ranking, presentation, and augmentation of product-specific metadata.
    \item[Collective knowledge] 
    Establishing connections with community initiatives aimed at uncovering and organizing metadata.
    While technology aggregates diverse sources, collaborative mechanisms~\citep{vanderSluis2022ARetrieval} can balance and weave together differing viewpoints and extract hidden information from nontraditional sources.
\end{description}

Although sustainability and other secondary information exist, they are often not available at the needed scale or in accessible formats. Product-level sustainability ratings, such as those from GoodOnYou (used by \citet{Cossatin2023}), are typically proprietary, while initiatives like EPREL, Blue Angel, and Higg Sustainability Profiles remain small, product-specific, and often not openly accessible or machine-readable \citep{GreenDB2022}. As a result, researchers often collect their own data and compute impact scores \citep{Cossatin2024,Tomkins2018}, though scaling these efforts remains difficult.
Alternative approaches are emerging. GreenDB, for instance, uses a web-scraping pipeline to provide a large-scale, continuously updated, machine-readable resource for clothing sustainability \citep{GreenDB2022}, while large language models have been applied to extract sustainability ratings from unstructured data \citep{Lin2023,Altammami2024}.
Crowdsourcing platforms also show promise. 
OpenFoodFacts (used in \citet{chazelas2020openfoodfacts}) 
exemplifies how community-driven efforts can aggregate and curate sustainability metadata.
However, comparable platforms for non-food products currently remain absent, despite active sustainability discussions on platforms like Reddit (e.g., \textit{r/sustainability}).
Together, these approaches point to promising directions for information extraction on secondary aspects, though significant challenges remain in achieving the scale and granularity required.

By framing responsible consumption as an information extraction problem, studies consider both diverse sources and the methods used to extract data. 
Although a substantial amount of information is available, and more is being generated through initiatives like the EU Corporate Sustainability Reporting Directive (CSRD), much of it is not machine-readable or easily understood by humans. 
The field of IR can take a leading role by providing the technology and space for different sources to converge \citep{vanderSluis2022ARetrieval}. 
Large language models (LLMs) hold particular promise in this effort, as they can parse, integrate, and rewrite disparate sources on secondary aspects into accessible formats. 
Moreover, as some critical information remains hidden, socio-technical solutions may be needed to extract those from nontraditional sources.
Such solutions offer a democratic and inclusive approach to integrating information, ensuring diverse organizations and voices are represented \citep{vandersluis2024jasist}.
Empowering community efforts has the potential to counterbalance more readily available marketing and sales-driven narratives.

\subsection{A Complex Search Task}\label{sec:complex}
Searching and shopping responsibly involves considering a wider range of criteria during the decision making process. These criteria include the typically product evaluation (primary) aspects along with a variety of ethical (secondary) aspects (Section \ref{challenges}, \textit{\nameref{cha:decision}}).
Similar to primary aspects, such secondary aspects demand domain-specific knowledge related to supply chains, materials, production locations, and domain-specific certifications to formulate appropriate search strategies and make responsible choices.
For instance, to check if clothing is sustainably made requires different knowledge  compared to assessing fair employment conditions for household goods.
Moreover, these aspects are even product-specific, meaning that for each product specific searches need to be carried out.
This will often require highly effortful and cognitively taxing searches~\cite{Schleenbecker2015InformationCoffee,Zander2012InformationFood} which 
necessitate navigating and browsing through a large variety of websites.
This challenge is compounded by the potential absence~\citep{Akerlof1970TheMechanism,Jacoby1974BrandLoad,Uusitalo2004EthicalFinland,Wiederhold2018EthicalIndustry} or difficulty to find or access \cite{lawrence2000accessibility,azzopardi2008accessibility} information due to information asymmetries, which can be particularly frustrating and may preclude search all together ~\citep{Brynjolfsson1999FrictionlessRetailers} (Section \ref{challenges}, \textit{\nameref{cha:search}}).
The extent of the challenges involved in incorporating secondary aspects into the decision space suggests that responsible consumerism necessitates solving a complex search task.
Such tasks are multifaceted, involving multiple information needs addressed by diverse sources, often spanning a number of search sessions. 
Moreover, they often demand in-depth domain knowledge to explore the different facets, requiring searchers to adopt new search strategies to find relevant and reliable information~\cite{Wildemuth2014UntanglingStudies, Liu2020IdentifyingTasks,Choi2019TheComplexity}.
The complexities are compounded when searchers lack the necessary task and domain knowledge to determine the next steps in their search~\cite{Awadallah2014SupportingTasks}. 
In sum, performing such complex search tasks means searchers need to synthesize an array of disparate, and potentially conflicting, information, leading to longer search durations, increased cognitive load, in order to make sense and make a purchasing decision (or not).
While mitigation strategies like Fair Trade and Better Cotton Initiative labeling can reduce some of the burden, many products lack such labels, and consumers may distrust or not fully understand the certification standards~\cite{Keller1987EffectsEffectiveness,Brynjolfsson1999FrictionlessRetailers}.

Given 
the information-seeking and decision-making challenges faced by responsible consumers,
we propose to address responsible consumerism as a complex search task. 
With this framing, the primary research focus for the \IR community should be directed towards reducing the complexity and making the task easier for consumers:
\begin{description}
    \item[Ranking signals \& retrieval performance] Due to information asymmetries, retrieval on secondary aspects is principally challenging. 
    Effective IR techniques can uncover hidden, obscured, or covered up information so that fair ranking systems \citep{raj2022dragons_princesses} can deliver a balanced and diversified representation. Ultimately, this calls for evaluating the precision, recall, and biases of IR systems on secondary aspects.
    \item[Search user interfaces (SUIs)] can limit the cognitive load users experience during this complex search task. 
    Effective SUIs, including conversational search assistants \citep{spatharioti2023}, can support consumers searching and navigating the decision space \citep{vandersluis2010}, and allow for information integration techniques that offer a comprehensive overview of the available options~\citep{Choi2021OrgBox:Tasks,Crescenzi2021SupportingOrgBox}. 
\end{description}

Search systems already play an important role in purchase decisions. 
IR systems support assessment of primary product aspects through features such as faceted navigation, advanced filtering, sorting, and product comparisons \citep{Hearst2006DesignInterfaces}.
An example of this within responsible consumption is Ecosia’s Green Consumption Assistant, 
which integrates eco-scores and sustainability tags in product comparisons \citep{bergener_einfluss_2023,bergener_increasing_2023}.
Substantial efforts have furthermore been underway towards sense-making support (search-as-learning, SAL; \citep{kamenihomte2022sal}), retrieval fairness, and sustainability-aware recommendations \citep{patro2020}. 
Sense-making can be supported by balancing textual complexity near a `sweet spot' of interest~\citep{vandersluis-2014-complexity}.
Fairness in IR ensures a wide range of perspectives rather than a market-driven dominance \citep{fang2024fairness}. 
While these algorithms demonstrate potential in retrieving understandable information about underrepresented aspects, traditional search engines often fall short in incorporating secondary details about products, resulting in commercially-biased SERPs \citep{Haider2022} and unmet needs for socially responsible consumers. 
Most interventions occur at the interface level, and retrieval experiments using secondary aspects as ranking signals remain scarce.

When considering responsible consumption as a complex search task, 
the goal becomes to support and enhance sense-making and decision-making. 
On the retrieval side, this includes enhancing the retrieval effectiveness of information on secondary aspects. 
Large-scale retrieval experiments (e.g., TREC challenges) can assess whether algorithms overcome biases to fairly represent hard-to-retrieve and otherwise unbalanced information on products. 
On the user side, this includes supporting users' sense-making of various secondary aspects for each product considered and achieving an overview of these aspects across products for decision making.
Here, cognitive and economic models~\citep{azzopardi2014economic,pirolli2003snifact} can guide interface designs, while eye-tracking, behavioral traces, and subjective feedback can reveal users' actual explorations and their sense of ease~\cite{buscher2009eye_tracking,cole2011reading_patterns,huang2012gaze}.
By considering both retrieval performance and user effectiveness, integrated solutions can be developed to make considering secondary aspects an easier task.

\subsection{A Knowledge Calibration Process}\label{sec:calibration}
In consumer decision-making, individuals often rely on incomplete information when evaluating products and services. Consumers tend to disregard missing information as well as form strong judgments based on the limited data available~\cite{Kardes2006DebiasingNeglect}. This makes consumers often overconfident of their knowledge, thinking they know more than they actually do~\cite{alba2000}.
This phenomenon is more pronounced in the context of responsible consumer decision-making. 
Individuals commonly make assumptions about various ethical, environmental, and social dimensions in the supply chain of products.
These assumptions that consumers make may not always be justified, especially considering the often intricate and extensive supply chains associated with the production of goods. 
Despite various campaigns drawing attention to different aspects such as tagging products with sustainability and environmental impact labels, it is easy to forget or neglect certain aspects, assume their irrelevance to the product category under consideration, and assume one's knowledge is sufficient (Section \ref{challenges}, \textit{\nameref{cha:awareness}}).

The challenges highlighted, including a lack of awareness regarding secondary aspects, their relevance, and potential knowledge gaps, point to a potential deficiency in knowledge calibration.
Knowledge calibration describes the alignment of (un)certainty with the accuracy and completeness of our knowledge~\citep{alba2000}. 
Ideally, we are certain when our knowledge is accurate and complete, and uncertain when it is inaccurate or incomplete~\citep{tormala2016}.
Uncertainty frequently drives information search activities aimed at restoring certainty~\cite{belkin1982,vandersluis2025ipm}, 
while
certainty about the accuracy of one's knowledge may deter further information acquisition~\cite{browne2007cognitive}.
Conversely, a sense of certainty regarding the completeness of one's knowledge can result in unawareness or overlooking aspects (e.g., over-confidence bias characteristic of the Dunning–Kruger effect~\cite{dunning2011dunning}). 
Search systems play an important role in this calibration process by shaping the judgmental context through the ranking and presentation of product information.
They provide knowledge context \citep{Smith2019Knowledge-ContextActions} that can either offer a sense of certainty via product rankings and metadata, like price and review scores, or highlight missing information, thereby increasing the salience of overlooked aspects~\citep{Kardes2006DebiasingNeglect}.
Search results on secondary aspects, by drawing attention to other decision variables~\citep{sanbonmatsu2003} and highlighting potential issues, can raise uncertainty regarding the consumer's knowledge. 
In fact, the search process can substantially shape purchase considerations even for consumers with low initial (EMCB) intentions~\citep{vandersluis2025chiir},
underlining the potential influence of search systems in supporting knowledge calibration.

We propose a shift in perspective by re-imagining search as a process of knowledge calibration. 
This shift directs our focus away from evaluating relevance or learning outcomes in interactive \IR studies and towards evaluating users' awareness of knowledge gaps and their subsequent seeking motivation and activity:
\begin{description}
    \item[Aspects awareness] Evaluating product searches on the range of aspects considered and searched for.
    By making secondary aspects a key metric of interactive evaluations, information systems are evaluated on their ability to contribute to awareness during product (information) searches \citep{sanbonmatsu2003}.
\item[Knowledge context] Expanding users' awareness and calibrating their knowledge of responsible aspects~\cite{sanbonmatsu2003}. 
Appropriate knowledge context can highlight overlooked or missing information, elucidate complexities when necessary, and support users in making more informed decisions~\citep{Smith2019Knowledge-ContextActions}.
\end{description}

Interactive IR studies have explored calibrating user knowledge through labels and prompts that highlight inaccuracies and incompleteness of search results \citep{rucker2014,vanderSluis2022ARetrieval}. These interface elements can reduce search time \citep{yamamoto2016credibility} and guide query behavior \citep{yamamoto2018querypriming}.
In addition, elements that highlight ethical, social, and environmental aspects hold promise for raising awareness of knowledge gaps~\citep{vandersluis2025ipm} and of missing dimensions in decision making \citep{Kardes2006DebiasingNeglect}. 
Effects may be moderated by factors like ranking position \citep{haas2017,unkel2017}, suggesting that interleaving results with secondary information (i.e., alongside primary product information) can enhance exposure. Although most studies have not targeted responsible consumption directly, similar nudging strategies in web shops -- using tags, prompts, product swap suggestions, and default options \citep{Islam2023,Cossatin2024,Tran2024} -- demonstrate the potential of interface elements to raise awareness and persuade consumers. 
Further research is needed to investigate how such interventions can shape product (information) searches.

Studies adopting a knowledge calibration perspective evaluate the cognitive effects \citep{joho2009cognitiveeffects} of various interventions in search interfaces and results rankings. 
In consumer psychology, these cognitive effects are measured as omission neglect \citep{Pfeiffer2014EffectsEvaluation} and knowledge confidence \citep{tormala2016}. Users' awareness can be evaluated by having participants list considered aspects, rate their feelings of knowing, and tracking search activity related to these aspects \citep{sanbonmatsu2003,vandersluis2025ipm}.
The application of these methodologies becomes feasible in the domain of responsible consumerism, where normative expectations pertain to the extent and accuracy with which responsible aspects should ideally be considered.
Nevertheless, there may be no single ``\textit{objectively optimal}'' decision outcome for this complex search task given the intricacies and trade-offs inherent in the decision-making process. %
Ideally, studies evaluate whether they can raise awareness and highlight knowledge gaps -- rather than change decision outcomes directly --
in order to help users develop and explore their preferences more actively \citep{knijnenburg2016}.

\section{Conclusion}
In this paper, we have outlined the information seeking challenges consumers face when trying to find, assess and decide among products when ethical, environmental and 
social aspects are of concern. 
Our survey results indicated that roughly two-thirds of participants were aware of these aspects, while another third cared about them. However, this awareness and desire to be more responsible when shopping often failed to translate into practice. 
To overcome this intention-behavior gap we have argued that \textit{responsible consumerism} requires \textit{responsible search} -- where the systems and platforms that consumers use not only help to educate and raise awareness, but also enable consumers in finding and factoring in such information when making their purchasing decisions.
Ideally, search systems would empower consumers to articulate their values, explore a more diverse range of options, reflect more deeply on the trade-offs, and ultimately make informed, context-sensitive, and hopefully more responsible decisions \citep{knijnenburg2016,patro2020}.

However, the challenges that responsible consumers face are unlikely to be met by just one (technical) system but rather by system(s) and human collaborative efforts (socio and technical) coming together: providing the legislation, regulation, education and information needed to facilitate more informed and responsible purchasing decisions. This will require building partnerships between retailers, NGOs, governments, and technology providers to create a comprehensive responsible shopping ecosystem that promotes awareness and facilitates access to information. 

In this paper, we have argued that IR has a crucial role to play in this transformation — by developing tools, interfaces, and systems that empower consumers through responsible search. 
In order to achieve this transformation, we need to (1) reframe responsible consumption as an information extraction problem to reduce information asymmetries; (2) redefine product search as a complex task requiring interfaces to lower the cost and burden of responsible search; and (3) reimagine search as a process of knowledge calibration to help consumers bridge gaps in their awareness and understanding when making purchasing decisions.
This will require addressing a variety of research challenges ranging from improving access and retrieval of secondary aspects to supporting sense making and decision making.
By embracing these challenges, the IR community can drive meaningful progress, shaping search systems that are not only technologically advanced but also socially responsible. 
This sets a path forward for interdisciplinary collaboration aimed at fostering responsible consumerism through responsible search.

\section{Our Perspective}
This paper is written by a multi-disciplinary team of academics from computer science and humanities. We aim to provide a consumer-focused perspective, 
rather than the commercial perspective that has typically focused on engagement, conversions and sales.

\begin{acks}
This work has received funding from the European Union’s Horizon 2020 Research and Innovation program under the Marie Skłodowska-Curie grant agreement No. 860721 (DoSSIER).
\end{acks}

\bibliographystyle{ACM-Reference-Format}
\balance
\bibliography{references.bib}

@article{chazelas2020openfoodfacts,
  title={Food additives: distribution and co-occurrence in 126,000 food products of the French market},
  author={Chazelas, Eloi and Deschasaux, Mélanie and Srour, Bernard and Kesse-Guyot, Emmanuelle and Julia, Chantal and Alles, Benjamin and Druesne-Pecollo, Nathalie and Galan, Pilar and Hercberg, Serge and Latino-Martel, Paule and Esseddik, Younes and Szabo, Fabien and Slamich, Pierre and Gigandet, Stephane and Touvier, Mathilde},
  journal={Scientific Reports},
  volume={10},
  number={1},
  pages={3980},
  year={2020},
  publisher={Nature Publishing Group},
  doi={10.1038/s41598-020-60948-w},
  _url={https://doi.org/10.1038/s41598-020-60948-w}
}

@article{spaniol2024greenwashing,
title = {Defining greenwashing: A concept analysis},
author = {Spaniol, Matthew J. and Danilova-Jensen, Evita and Nielsen, Martin and Rosdahl, Carl Gyldenkaerne and Schmidt, Clara Jasmin},
pages = {9055},
_url = {https://www.mdpi.com/2071-1050/16/20/9055},
year = {2024},
month = {oct},
day = {19},
_urldate = {2025-04-24},
journal = {Sustainability},
volume = {16},
number = {20},
issn = {2071-1050},
doi = {10.3390/su16209055},
abstract = {The lack of a shared, operant definition for greenwashing has led to fragmented scholarly research, unclear guidelines for practice, inconsistent enforcement, and reactive policy frameworks; resulting in ineffective efforts to combat its growth. Using concept analysis, this research establishes a composite definition for greenwashing by identifying the constitutive attributes sourced across 79 scholarly definitions. The analysis finds six requirements necessary for identifying greenwashing: a claim on environmental performance by a private sector organization marketing a product or a service, which cannot be substantiated, made with deceptive intent, and done to establish a competitive advantage. Fulfilling these criteria warrants an accusation of greenwashing. With the aim to prevent its further spread and misuse, the article provides a diagnostic tool for separating similar but often conflated concepts from greenwashing to organize scholarly research, provide guidelines for practitioners, and support regulators\textquoteright case analysis.}
}

@inproceedings{vandersluis2025chiir,
  author       = {Frans van der Sluis and Leif Azzopardi},
  title        = {Search Changes Consumers’ Minds: How Recognizing Gaps in Understanding Drives Ethical Choices},
  booktitle    = {Proceedings of the 2025 ACM SIGIR Conference on Human Information Interaction and Retrieval (CHIIR '25)},
  year         = {2025},
  pages        = {1--14},
  doi          = {10.1145/3698204.3716456},
}

@inproceedings{azzopardi2008retrievability,
author = {Azzopardi, Leif and Vinay, Vishwa},
title = {Retrievability: an evaluation measure for higher order information access tasks},
year = {2008},
doi = {10.1145/1458082.1458157},
booktitle = {Proceedings of the 17th ACM Conference on Information and Knowledge Management},
pages = {561–570},
numpages = {10},
series = {CIKM '08}
}

@inproceedings{beiga2018equity,
author = {Biega, Asia J. and Gummadi, Krishna P. and Weikum, Gerhard},
title = {Equity of Attention: Amortizing Individual Fairness in Rankings},
year = {2018},
doi = {10.1145/3209978.3210063},
booktitle = {The 41st International ACM SIGIR Conference on Research \& Development in Information Retrieval},
pages = {405–414},
numpages = {10},
series = {SIGIR '18}
}

@inproceedings{white2013biases,
author = {White, Ryen},
title = {Beliefs and biases in web search},
year = {2013},
doi = {10.1145/2484028.2484053},
booktitle = {Proceedings of the 36th International ACM SIGIR Conference on Research and Development in Information Retrieval},
pages = {3–12},
numpages = {10},
series = {SIGIR '13}
}

@article{yang2020greenwashing,
  title={Greenwashing behaviours: Causes, taxonomy and consequences based on a systematic literature review},
  author={Yang, Zhi and Nguyen, Thi Thu Huong and Nguyen, Hoang Nam and Nguyen, Thi Thuy Nga and Cao, Thi Thanh},
  journal={Journal of business economics and management},
  volume={21},
  number={5},
  pages={1486--1507},
  year={2020},
doi= {https://doi.org/10.3846/jbem.2020.13225}
}

@article{shchory2020information,
  title={Information asymmetries in e-commerce: The challenge of credence qualities},
  author={Shchory, Noga Blickstein},
  journal={Journal of High Technology Law},
  volume={20},
  pages={1},
  year={2020},
  publisher={SSRN}
}

@InProceedings{wilkie2014bias,
author="Wilkie, Colin
and Azzopardi, Leif",
editor="de Rijke, Maarten
and Kenter, Tom
and de Vries, Arjen P.
and Zhai, ChengXiang
and de Jong, Franciska
and Radinsky, Kira
and Hofmann, Katja",
title="Best and Fairest: An Empirical Analysis of Retrieval System Bias",
booktitle="Advances in Information Retrieval",
year="2014",
doi = "https://doi.org/10.1007/978-3-319-06028-6_2",
publisher="Springer International Publishing",
address="Cham",
pages="13--25",
series = {ECIR `14},
isbn="978-3-319-06028-6"
}

@inproceedings{patro2020,
title = {Towards Safety and Sustainability: Designing Local Recommendations for Post-pandemic World},
author = {Patro, Gourab K and Chakraborty, Abhijnan and Banerjee, Ashmi and Ganguly, Niloy},
pages = {358-367},
doi = {10.1145/3383313.3412251},
year = {2020},
booktitle = {Fourteenth {ACM} Conference on Recommender Systems},
series = {RecSys `20}
}

@inproceedings{knijnenburg2016,
title = {Recommender Systems for Self-Actualization},
author = {Knijnenburg, Bart P. and Sivakumar, Saadhika and Wilkinson, Daricia},
pages = {11-14},
publisher = {ACM},
url = {https://dl.acm.org/doi/10.1145/2959100.2959189},
year = {2016},
month = {sep},
day = {7},
urldate = {2025-04-16},
isbn = {9781450340359},
doi = {10.1145/2959100.2959189},
address = {New York, {NY}, {USA}},
booktitle = {Proceedings of the 10th {ACM} Conference on Recommender Systems},
series = {RecSys '16}
}

@article{unkel2017,
title = {The effects of credibility cues on the selection of search engine results},
author = {Unkel, Julian and Haas, Alexander},
pages = {1850-1862},
url = {http://doi.wiley.com/10.1002/asi.23820},
year = {2017},
month = {aug},
urldate = {2018-03-01},
journal = {Journal of the Association for Information Science and Technology},
volume = {68},
number = {8},
issn = {23301635},
doi = {10.1002/asi.23820}
}

@article{haas2017,
title = {Ranking versus reputation: perception and effects of search result credibility},
author = {Haas, Alexander and Unkel, Julian},
pages = {1285-1298},
url = {https://www.tandfonline.com/doi/full/10.1080/0144929X.2017.1381166},
year = {2017},
month = {12},
day = {2},
urldate = {2021-02-11},
journal = {Behaviour \& information technology},
volume = {36},
number = {12},
issn = {0144-{929X}},
doi = {10.1080/0144929X.2017.1381166}
}

@article{kamenihomte2022sal,
title = {Search engines in learning contexts: A literature review},
author = {Kameni Homte, Jaurès Styve and Batchakui, Bernabé and Nkambou, Roger},
pages = {254-272},
url = {https://online-journals.org/index.php/i-jet/article/view/26217},
year = {2022},
month = {jan},
day = {31},
urldate = {2025-02-19},
journal = {International Journal of Emerging Technologies in Learning ({iJET})},
volume = {17},
number = {02},
issn = {1863-0383},
doi = {10.3991/ijet.v17i02.26217}
}

@article{spatharioti2023,
title = {Comparing Traditional and {LLM}-based Search for Consumer Choice: A Randomized Experiment},
author = {Spatharioti, Sofia Eleni and Rothschild, David M. and Goldstein, Daniel G. and Hofman, Jake M.},
url = {https://arxiv.org/abs/2307.03744},
year = {2023},
urldate = {2025-02-18},
journal = {arXiv},
doi = {10.48550/arxiv.2307.03744}
}

@inproceedings{azzopardi2014economic,
author = {Azzopardi, Leif},
title = {Modelling interaction with economic models of search},
year = {2014},
doi = {10.1145/2600428.2609574},
booktitle = {Proceedings of the 37th International ACM SIGIR Conference on Research \& Development in Information Retrieval},
pages = {3–12},
numpages = {10},
series = {SIGIR '14}
}

@inproceedings{pirolli2003snifact,
author = {Pirolli, Peter and Fu, Wai-tat},
year = {2003},
month = {03},
pages = {},
title = {SNIF-ACT: A model of information foraging on the world wide web},
isbn = {978-3-540-40381-4},
journal = {Proceedings of the 9th International Conference on User Modeling},
series = {UMAP `03},
doi = {10.1007/3-540-44963-9_8}
}

@inproceedings{huang2012gaze,
author = {Huang, Jeff and White, Ryen and Buscher, Georg},
title = {User see, user point: gaze and cursor alignment in web search},
year = {2012},
_isbn = {9781450310154},
_publisher = {Association for Computing Machinery},
_address = {New York, NY, USA},
_url = {https://doi.org/10.1145/2207676.2208591},
doi = {10.1145/2207676.2208591},
abstract = {Past studies of user behavior in Web search have correlated eye-gaze and mouse cursor positions, and other lines of research have found cursor interactions to be useful in determining user intent and relevant parts of Web pages. However, cursor interactions are not all the same; different types of cursor behavior patterns exist, such as reading, hesitating, scrolling and clicking, each of which has a different meaning. We conduct a search study with 36 subjects and 32 search tasks to determine when gaze and cursor are aligned, and thus when the cursor position is a good proxy for gaze position. We study the effect of time, behavior patterns, user, and search task on the gaze-cursor alignment, findings which lead us to question the maxim that "gaze is well approximated by cursor." These lessons inform an experiment in which we predict the gaze position with better accuracy than simply using the cursor position, improving the state-of-the-art technique for approximating visual attention with the cursor. Our new technique can help make better use of large-scale cursor data in identifying how users examine Web search pages.},
booktitle = {Proceedings of the SIGCHI Conference on Human Factors in Computing Systems},
pages = {1341–1350},
numpages = {10},
_keywords = {search examination behavior, gaze, eye-tracking, cursor},
_location = {Austin, Texas, USA},
series = {CHI '12}
}

@article{cole2011reading_patterns,
    author = {Cole, Michael J. and Gwizdka, Jacek and Liu, Chang and Bierig, Ralf and Belkin, Nicholas J. and Zhang, Xiangmin},
    title = {Task and user effects on reading patterns in information search},
    journal = {Interacting with Computers},
    volume = {23},
    number = {4},
    pages = {346-362},
    year = {2011},
    month = {05},
}

@article{vandersluis-2014-complexity,
  author    = {Frans van der Sluis and
               Egon L. van den Broek and
               Richard Glassey and
               Elisabeth M. A. G. van Dijk and
               Franciska M. G. de Jong},
  title     = {When complexity becomes interesting},
  journal   = {Journal of the Association for Information Science and Technology},
  volume    = {65},
  number    = {7},
  pages     = {1478--1500},
  year      = {2014},
  url       = {https://doi.org/10.1002/asi.23095},
  doi       = {10.1002/asi.23095},
  bibsource = {dblp computer science bibliography, https://dblp.org}
}

@inproceedings{yamamoto2016credibility,
author = {Yamamoto, Yusuke and Shimada, Satoshi},
title = {Can Disputed Topic Suggestion Enhance User Consideration of Information Credibility in Web Search?},
year = {2016},
booktitle = {Proceedings of the 27th ACM Conference on Hypertext and Social Media},
pages = {169–177},
numpages = {9},
doi = {https://doi.org/10.1145/2914586.2914592},
_keywords = {web search, decision making, credibility, careful information seeking, behavior analysis},
_location = {Halifax, Nova Scotia, Canada},
series = {HT '16}
}

@inproceedings{buscher2009eye_tracking,
author = {Buscher, Georg and Cutrell, Edward and Morris, Meredith Ringel},
title = {What do you see when you're surfing? using eye tracking to predict salient regions of web pages},
year = {2009},
isbn = {9781605582467},
publisher = {Association for Computing Machinery},
address = {New York, NY, USA},
url = {https://doi.org/10.1145/1518701.1518705},
doi = {10.1145/1518701.1518705},
booktitle = {Proceedings of the SIGCHI Conference on Human Factors in Computing Systems},
pages = {21–30},
numpages = {10},
location = {Boston, MA, USA},
series = {CHI '09}
}

@article{fang2024fairness,
title = {Fairness in search systems},
author = {Fang, Yi and Singh, Ashudeep and Tao, Zhiqiang},
pages = {262-416},
url = {http://www.nowpublishers.com/article/Details/{INR}-101},
year = {2024},
urldate = {2025-01-29},
journal = {Foundations and Trends\textregistered in Information Retrieval},
volume = {18},
number = {3},
issn = {1554-0669},
doi = {10.1561/1500000101}
}

@article{vandersluis2025ipm,
title = {Wanting information: Uncertainty and its reduction through search engagement},
author = {van der Sluis, Frans},
pages = {103890},
year = {2025},
journal = {Information Processing \& Management},
volume = {62},
number = {2},
doi = {10.1016/j.ipm.2024.103890}
}

@inproceedings{joho2009cognitiveeffects,
  author    = {Hideo Joho},
  title     = {Cognitive Effects in Information Seeking and Retrieval},
  booktitle = {ECIR '09 Workshop on Contextual Information Access, Seeking and Retrieval Evaluation (CIRSE '09)},
  year      = {2009},
  address   = {Toulouse, France},
  month     = apr,
  urldate   = {2019-05-02}
}

@inproceedings{yamamoto2018querypriming,
title = {Query priming for promoting critical thinking in web search},
author = {Yamamoto, Yusuke and Yamamoto, Takehiro},
series = {{CHIIR} '18},
pages = {12-21},
year = {2018},
doi = {10.1145/3176349.3176377},
booktitle = {Proceedings of the 2018 Conference on Human Information Interaction and Retrieval}
}

@inproceedings{Tran2024,
author = {Tran, Thi Ngoc Trang and Polat Erdeniz, Seda and Felfernig, Alexander and Lubos, Sebastian and El Mansi, Merfat and Le, Viet-Man},
title = {Less is More: Towards Sustainability-Aware Persuasive Explanations in Recommender Systems},
year = {2024},
_isbn = {9798400705052},
_publisher = {Association for Computing Machinery},
_address = {New York, NY, USA},
_url = {https://doi.org/10.1145/3640457.3691708},
doi = {10.1145/3640457.3691708},
booktitle = {Proceedings of the 18th ACM Conference on Recommender Systems},
pages = {1108–1112},
numpages = {5},
_location = {Bari, Italy},
series = {RecSys '24}
}

@techreport{bergener_einfluss_2023,
	title = {Der {Einfluss} von {Nachhaltigkeitshinweisen} auf {Kaufentscheidungen}\#05 {Driving} sustainable choices for consumer electronics: the influence of sustainability cues on purchasing decisions},
	Z = {Creative Commons Attribution 4.0 International},
	url = {https://depositonce.tu-berlin.de/handle/11303/20390},
	doi = {10.14279/depositonce-19188},
    institution = {Technische Universität Berlin},
	language = {en},
	author = {Bergener, Jens and Jadkowski, Robin and Korenke, Ruben and Jankowski, Patricia and Veneny, Marek},
	collaborator = {{Technische Universität Berlin}},
	year = {2023},
}

@techreport{bergener_increasing_2023,
	title = {Increasing the effectiveness of sustainability tags. {Empirical} evidence from user tests and a field study on {Ecosia} {Shopping}.},
	copyright = {Creative Commons Attribution 4.0 International},
	url = {https://depositonce.tu-berlin.de/handle/11303/20891},
	doi = {10.14279/depositonce-19689},
    institution = {Technische Universität Berlin},
	language = {en},
	urldate = {2025-01-30},
	author = {Bergener, Jens and Gossen, Maike and Veneny, Marek and Paszternak, Zsofia and Korenke, Ruben},
	collaborator = {{Technische Universität Berlin}},
	year = {2023},
}

@inproceedings{Lin2023,
  title     = {SUSTAINABLESIGNALS: An AI Approach for Inferring Consumer Product Sustainability},
  author    = {Lin, Tong and Xu, Tianliang and Zac, Amit and Tomkins, Sabina},
  booktitle = {Proceedings of the Thirty-Second International Joint Conference on Artificial Intelligence},
  _publisher = {International Joint Conferences on Artificial Intelligence Organization},
series = {{IJCAI `23}},
  _editor    = {Edith Elkind},
  pages     = {6067--6075},
  year      = {2023},
  _month     = {8},
  _note      = {AI for Good},
  doi       = {10.24963/ijcai.2023/673},
  _url       = {https://doi.org/10.24963/ijcai.2023/673},
}

@inproceedings{Altammami2024,
author = {Altammami, Alaa and Dimitrova, Vania and Pournaras, Evangelos},
title = {What you see, What you get? Mapping Inconsistencies of Sustainability Judgements among Experts and Consumers},
year = {2024},
doi = {10.1145/3677525.3678695},
booktitle = {Proceedings of the 2024 International Conference on Information Technology for Social Good},
pages = {443–452},
numpages = {10},
location = {Bremen, Germany},
series = {GoodIT '24}
}

@misc{GreenDB2022,
title={GreenDB -- A Dataset and Benchmark for Extraction of Sustainability Information of Consumer Goods}, 
author={Sebastian Jäger and Alexander Flick and Jessica Adriana Sanchez Garcia and Kaspar von den Driesch and Karl Brendel and Felix Biessmann},
year={2022},
eprint={2207.10733},
archivePrefix={arXiv},
primaryClass={cs.LG},
url={https://arxiv.org/abs/2207.10733}, 
}

@inproceedings{Tomkins2018,
author = {Tomkins, Sabina and Isley, Steven and London, Ben and Getoor, Lise},
title = {Sustainability at scale: towards bridging the intention-behavior gap with sustainable recommendations},
year = {2018},
_isbn = {9781450359016},
_publisher = {Association for Computing Machinery},
_address = {New York, NY, USA},
_url = {https://doi.org/10.1145/3240323.3240411},
doi = {10.1145/3240323.3240411},
booktitle = {Proceedings of the 12th ACM Conference on Recommender Systems},
pages = {214–218},
numpages = {5},
_keywords = {sustainability, probabilistic programming},
_location = {Vancouver, British Columbia, Canada},
series = {RecSys '18}
}

@article{Islam2023,
title = {SEER: Sustainable E-commerce with Environmental-impact Rating},
journal = {Cleaner Environmental Systems},
volume = {8},
pages = {100104},
year = {2023},
_issn = {2666-7894},
doi = {https://doi.org/10.1016/j.cesys.2022.100104},
_url = {https://www.sciencedirect.com/science/article/pii/S2666789422000356},
author = {Md Saiful Islam and Adiba Mahbub Proma and Caleb Wohn and Karen Berger and Serena Uong and Varun Kumar and Katrina Smith Korfmacher and Ehsan Hoque}}

@INPROCEEDINGS{Cossatin2023,
  author={Cossatin, Angelo Geninatti and Mauro, Noemi and Ardissono, Liliana},
  booktitle={IEEE/WIC International Conference on Web Intelligence and Intelligent Agent Technology}, 
  series={WI-IAT `23},
  _location={Venice, Italy},
  _address={Los Alamitos, CA, USA},
  title={Enriching Recommender Systems Results with Data about Sustainability and Ethical Standards of Brands}, 
  year={2023},
  pages={238-242},
  keywords={Ethics;Green products;User interfaces;Behavioral sciences;Synchronization;Intelligent agents;Sustainable development;multi-list user interfaces;sustainability;fashion;human-centric computing and services},
  doi={10.1109/WI-IAT59888.2023.00037}
}

@ARTICLE{Cossatin2024,
  author={Cossatin, Angelo Geninatti and Mauro, Noemi and Ardissono, Liliana},
  journal={IEEE Access}, 
  title={Promoting Green Fashion Consumption Through Digital Nudges in Recommender Systems}, 
  year={2024},
  volume={12},
  number={},
  pages={6812-6829},
  doi={10.1109/ACCESS.2024.3349710}}

@misc{zhou2024,
      title={Advancing Sustainability via Recommender Systems: A Survey}, 
      author={Xin Zhou and Lei Zhang and Honglei Zhang and Yixin Zhang and Xiaoxiong Zhang and Jie Zhang and Zhiqi Shen},
      year={2024},
      _eprint={2411.07658},
      publisher={arXiv},
      _primaryClass={cs.IR},
      doi={https://doi.org/10.48550/arXiv.2411.07658}, 
}

@inproceedings{Scells2022,
author = {Scells, Harrisen and Zhuang, Shengyao and Zuccon, Guido},
title = {Reduce, Reuse, Recycle: Green Information Retrieval Research},
year = {2022},
_isbn = {9781450387323},
_publisher = {Association for Computing Machinery},
_address = {New York, NY, USA},
_url = {https://doi.org/10.1145/3477495.3531766},
doi = {10.1145/3477495.3531766},
booktitle = {Proceedings of the 45th International ACM SIGIR Conference on Research and Development in Information Retrieval},
pages = {2825–2837},
numpages = {13},
keywords = {green ir, emissions, efficiency, deep learning},
location = {Madrid, Spain},
series = {SIGIR '22}
}

@inproceedings{Vente2024,
author = {Vente, Tobias and Wegmeth, Lukas and Said, Alan and Beel, Joeran},
title = {From Clicks to Carbon: The Environmental Toll of Recommender Systems},
year = {2024},
_isbn = {9798400705052},
_publisher = {Association for Computing Machinery},
_address = {New York, NY, USA},
_url = {https://doi.org/10.1145/3640457.3688074},
doi = {10.1145/3640457.3688074},
booktitle = {Proceedings of the 18th ACM Conference on Recommender Systems},
pages = {580–590},
numpages = {11},
_location = {Bari, Italy},
series = {RecSys '24}
}

@inproceedings{Spillo2023,
author = {Spillo, Giuseppe and De Filippo, Allegra and Musto, Cataldo and Milano, Michela and Semeraro, Giovanni},
title = {Towards Sustainability-aware Recommender Systems: Analyzing the Trade-off Between Algorithms Performance and Carbon Footprint},
year = {2023},
_isbn = {9798400702419},
_publisher = {Association for Computing Machinery},
_address = {New York, NY, USA},
_url = {https://doi.org/10.1145/3604915.3608840},
doi = {10.1145/3604915.3608840},
booktitle = {Proceedings of the 17th ACM Conference on Recommender Systems},
pages = {856–862},
numpages = {7},
_keywords = {carbon footprint, evaluation, non-accuracy metrics, recommender systems, sustainability},
_location = {Singapore, Singapore},
series = {RecSys '23}
}

@ARTICLE{Felfernig2023,
AUTHOR={Felfernig, Alexander  and Wundara, Manfred  and Tran, Thi Ngoc Trang  and Polat-Erdeniz, Seda  and Lubos, Sebastian  and El Mansi, Merfat  and Garber, Damian  and Le, Viet-Man },
TITLE={Recommender systems for sustainability: overview and research issues},
JOURNAL={Frontiers in Big Data},
VOLUME={6},
YEAR={2023},
URL={https://www.frontiersin.org/journals/big-data/articles/10.3389/fdata.2023.1284511},
DOI={10.3389/fdata.2023.1284511}}

@misc{jannach2024rs4good,
      title={Recommender Systems for Good (RS4Good): Survey of Use Cases and a Call to Action for Research that Matters}, 
      author={Dietmar Jannach and Alan Said and Marko Tkalčič and Markus Zanker},
      year={2024},
      eprint={2411.16645},
      archivePrefix={arXiv},
      primaryClass={cs.IR},
      url={https://arxiv.org/abs/2411.16645}, 
}

@article{Mager2012,
author = {Astrid Mager},
title = {Algorithmic Ideology},
journal = {Information, Communication \& Society},
volume = {15},
number = {5},
pages = {769--787},
year = {2012},
publisher = {Routledge},
doi = {10.1080/1369118X.2012.676056},
_URL = {https://doi.org/10.1080/1369118X.2012.676056},
_eprint = {https://doi.org/10.1080/1369118X.2012.676056}
}

@article{Haider2022,
title = {Algorithmically embodied emissions: the environmental harm of everyday life information in digital culture},
  author    = {Haider, J. and Rödl, M. and Joosse, S.},
url = {http://informationr.net/ir/27-{SpIssue}/{CoLIS2022}/colis2224.html},
year = {2022},
urldate = {2025-04-16},
journal = {Information Research: an international electronic journal},
volume = {27},
issn = {13681613},
doi = {10.47989/colis2224}
}

@article{Gleason2024, title={Search Engine Revenue from Navigational and Brand Advertising}, volume={18}, url={https://ojs.aaai.org/index.php/ICWSM/article/view/31329}, DOI={10.1609/icwsm.v18i1.31329}, abstractNote={Keyword advertising on general web search engines is a multi-billion dollar business. Keyword advertising turns contentious, however, when businesses target ads against their competitors’ brand names---a practice known as &quot;competitive poaching.&quot; To stave off poaching, companies defensively bid on ads for their own brand names. Google, in particular, has faced lawsuits and regulatory scrutiny since it altered its policies in 2004 to allow poaching. In this study, we investigate the sources of advertising revenue earned by Google, Bing, and DuckDuckGo by examining ad impressions, clicks, and revenue on navigational and brand searches. Using logs of searches performed by a representative panel of US residents, we estimate that ads on these searches account for 28--36% of Google’s search revenue, while Bing earns even more. We also find that the effectiveness of these ads for advertisers varies. We conclude by discussing the implications of our findings for advertisers and regulators.}, number={1}, journal={Proceedings of the International AAAI Conference on Web and Social Media}, author={Gleason, Jeffrey and Koeninger, Alice and Hu, Desheng and Teurn, Jessica and Bart, Yakov and Knight, Samsun and Robertson, Ronald E. and Wilson, Christo}, year={2024}, month={May}, pages={488-501} }

@article{rucker2014,
title = {Consumer conviction and commitment: An appraisal-based framework for attitude certainty},
author = {Rucker, Derek D. and Tormala, Zakary L. and Petty, Richard E. and Briñol, Pablo},
pages = {119-136},
_url = {http://doi.wiley.com/10.1016/j.jcps.2013.07.001},
year = {2014},
month = {jan},
urldate = {2021-02-11},
journal = {Journal of Consumer Psychology},
volume = {24},
number = {1},
issn = {10577408},
doi = {10.1016/j.jcps.2013.07.001}
}

@inproceedings{azzopardi2024src, 
author = {Azzopardi, Leif and Van Der Sluis, Frans}, 
title = {Seeking Socially Responsible Consumers: Exploring the Intention-Search-Behaviour Gap}, year = {2024}, 
doi = {10.1145/3627508.3638324}, booktitle = {Proceedings of the 2024 Conference on Human Information Interaction and Retrieval}, pages = {153–164},
numpages = {12}, 
series = {CHIIR '24} }

@article{vandersluis2024jasist,
title = {An empirical exploration of the subjectivity problem of information qualities},
author = {Van der Sluis, Frans and Faure, Julien and Homnual, Sofie Phutachard},
year = {2024},
journal = {Journal of the Association for Information Science and Technology},
doi = {10.1002/asi.24884}
}

@inproceedings{vandersluis2010,
title = {Information Retrieval {eXperience} ({IRX}): Towards a Human-Centered Personalized Model of Relevance},
author = {van der Sluis, Frans and van den Broek, Egon L. and van Dijk, Betsy},
pages = {322-325},
year = {2010},
series = {WI-IAT `10},
doi = {10.1109/WI-IAT.2010.222},
booktitle = {2010 {IEEE}/{WIC}/{ACM} International Conference on Web Intelligence and Intelligent Agent Technology}
}

@inproceedings{azzopardi2009greener_search,
author = {Azzopardi, Leif and Vanderbauwhede, Wim and Moadeli, Mahmoud},
title = {Developing energy efficient filtering systems},
year = {2009},
booktitle = {Proceedings of the 32nd International ACM SIGIR Conference on Research and Development in Information Retrieval},
pages = {664–665},
numpages = {2},
series = {SIGIR '09}
}

@InProceedings{azzopardi2008accessibility,
author="Azzopardi, Leif
and Vinay, Vishwa",
title="Accessibility in Information Retrieval",
doi="10.1007/978-3-540-78646-7_46",
booktitle="Advances in Information Retrieval",
year="2008",
pages="482--489",
}

@article{lawrence2000accessibility,
  author    = {Steve Lawrence and C. Lee Giles},
  title     = {Accessibility of information on the web},
  journal   = {Nature},
  volume    = {400},
  number    = {6740},
  pages     = {107--107},
  year      = {1999},
  doi       = {10.1038/21987}
}

@article{SUDBURYRILEY20162697,
title = {Ethically minded consumer behavior: Scale review, development, and validation},
journal = {Journal of Business Research},
volume = {69},
number = {8},
pages = {2697-2710},
year = {2016},
doi = {https://doi.org/10.1016/j.jbusres.2015.11.005},
author = {Lynn Sudbury-Riley and Florian Kohlbacher}
}

@article{carrigan2004better_shopping,
author = {Carrigan, Marylyn and Szmigin, Isabelle and Wright, Joanne},
year = {2004},
month = {10},
title = {Shopping for a better world? An interpretive study of the potential for ethical consumption within the older market},
volume = {21},
journal = {Journal of Consumer Marketing},
doi = {10.1108/07363760410558672}
}

@article{tormala2016,
title = {The role of certainty (and uncertainty) in attitudes and persuasion},
author = {Tormala, Zakary L},
pages = {6-11},
_url = {http://linkinghub.elsevier.com/retrieve/pii/{S2352250X1530004X}},
year = {2016},
_month = {aug},
urldate = {2020-08-12},
journal = {Current opinion in psychology},
volume = {10},
_issn = {2352250X},
doi = {10.1016/j.copsyc.2015.10.017},
abstract = {Psychological certainty plays a key role in shaping people\textquoterights thoughts, judgments, attitudes, and behaviors. This article provides an overview of recent work on attitude certainty, which has been the subject of considerable attention in the social and consumer psychology literatures. In particular, this article describes the consequences of feeling certain or uncertain of an attitude, outlines the metacognitive appraisals that shape people\textquoterights feelings of certainty or uncertainty, and highlights recent developments suggesting that strategically inducing uncertainty during message processing can enhance message impact. In essence, whereas uncertainty can stimulate processing and create a desire for information, certainty helps give an attitude durability and impact. Address}
}

@article{alba2000,
title = {Knowledge calibration: what consumers know and what they think they know},
author = {Alba, Joseph W. and Hutchinson, J. Wesley},
pages = {123-156},
_url = {https://academic.oup.com/jcr/article-lookup/doi/10.1086/314317},
year = {2000},
_month = {sep},
urldate = {2021-03-19},
journal = {The Journal of Consumer Research},
volume = {27},
number = {2},
_issn = {0093-5301},
doi = {10.1086/314317},
abstract = {Consumer knowledge is seldom complete or errorless. Therefore, the self-assessed validity of knowledge and consequent knowledge calibration (i.e., the correspondence between self-assessed and actual validity) is an important issue for the study of consumer decision making. In this article we describe methods and models used in calibration research. We then review a wide variety of empirical results indicating that high levels of calibration are achieved rarely, moderate levels that include some degree of systematic bias are the norm, and confidence and accuracy are sometimes completely uncorrelated. Finally, we examine the explanations of miscalibration and offer suggestions for future research.}
}

@inproceedings{raj2022dragons_princesses,
  author    = {Amifa Raj and Michael D. Ekstrand},
  title     = {Fire Dragon and Unicorn Princess: Gender Stereotypes and Children’s Products in Search Engine Responses},
  booktitle = {Proceedings of the ACM SIGIR Workshop on eCommerce},
series = {SIGIR eCom ’22},
  year      = {2022},
  _publisher = {ACM},
  _address   = {New York, NY, USA},
  pages     = {1--9},
doi = {https://doi.org/10.48550/arXiv.2206.13747},
  _note      = {9 pages}
}

@inproceedings{vanderSluis2022ARetrieval,
    title = {{A Conversationalist Approach to Information Quality in Information Interaction and Retrieval}},
    year = {2022},
    booktitle = {CHIIR '22 Workshop on Information Quality in Information Interaction and Retrieval},
    author = {van der Sluis, Frans},
    doi = {https://doi.org/10.48550/arXiv.2210.07296}
}

@article{Schmidt1996ASearch,
    title = {{A proposed model of external consumer information search}},
    year = {1996},
    journal = {Journal of the Academy of Marketing Science},
    author = {Schmidt, Jeffrey B. and Spreng, Richard A.},
    number = {3},
    pages = {246--256},
    volume = {24},
    publisher = {SAGE Publications Inc.},
    doi = {10.1177/0092070396243005},
    issn = {00920703}
}

@article{Broder2002ASearch,
author = {Broder, Andrei},
title = {A taxonomy of web search},
year = {2002},
issue_date = {Fall 2002},
publisher = {Association for Computing Machinery},
address = {New York, NY, USA},
volume = {36},
number = {2},
issn = {0163-5840},
url = {https://doi.org/10.1145/792550.792552},
doi = {10.1145/792550.792552},
journal = {SIGIR Forum},
month = sep,
pages = {3–10},
numpages = {8}
}

@article{Punj1983AnMaking,
    title = {{An Interaction Framework of Consumer Decision Making}},
    year = {1983},
    journal = {Journal of Consumer Research},
    author = {Punj, Girish N and Stewart, David W},
    number = {2},
    month = {9},
    pages = {181--196},
    volume = {10},
    url = {https://doi.org/10.1086/208958},
    doi = {10.1086/208958},
    issn = {0093-5301}
}

@article{Flavian-Blanco2011AnalyzingEngines,
    title = {{Analyzing the emotional outcomes of the online search behavior with search engines}},
    year = {2011},
    journal = {Computers in Human Behavior},
    author = {Flavi{\'{a}}n-Blanco, Carlos and Gurrea-Sarasa, Raquel and Or{\'{u}}s-Sanclemente, Carlos},
    number = {1},
    month = {1},
    pages = {540--551},
    volume = {27},
    doi = {10.1016/j.chb.2010.10.002},
    issn = {07475632}
}

@article{Jacoby1974BrandLoad,
author = {Jacob Jacoby and Donald E. Speller and Carol A. Kohn},
title ={Brand Choice Behavior as a Function of Information Load},
journal = {Journal of Marketing Research},
volume = {11},
number = {1},
pages = {63-69},
year = {1974},
doi = {10.1177/002224377401100106}
}

@article{Tsagkias2021ChallengesRecommendations,
    title = {{Challenges and Research Opportunities in ECommerce Search and Recommendations}},
    year = {2021},
    journal = {SIGIR Forum},
    author = {Tsagkias, Manos and King, Tracy Holloway and Kallumadi, Surya and Murdock, Vanessa and de Rijke, Maarten},
    number = {1},
    month = {2},
    volume = {54},
    _publisher = {Association for Computing Machinery},
    _url = {https://doi.org/10.1145/3451964.3451966},
    _address = {New York, NY, USA},
    doi = {10.1145/3451964.3451966},
    _issn = {0163-5840}
}

@article{browne2007cognitive,
 ISSN = {02767783},
 _URL = {http://www.jstor.org/stable/25148782},
doi = {10.2307/25148782},
 abstract = {Online search has become a significant activity in the daily lives of individuals throughout much of the world. The almost instantaneous availability of billions of web pages has caused a revolution in the way people seek information. Despite the increasing importance of online search behavior in decision making and problem solving, very little is known about why people stop searching for information online. In this paper, we review the literature concerning online search and cognitive stopping rules, and then describe specific types of information search tasks. Based on this theoretical development, we generated hypotheses and conducted an experiment with 115 participants each performing three search tasks on the web. Our findings show that people utilize a number of stopping rules to terminate search, and that the stopping rule used depends on the type of task performed. Implications for online information search theory and practice are discussed.},
 author = {Glenn J. Browne and Mitzi G. Pitts and James C. Wetherbe},
 journal = {MIS Quarterly},
 number = {1},
 pages = {89--104},
 publisher = {Management Information Systems Research Center, University of Minnesota},
 title = {Cognitive Stopping Rules for Terminating Information Search in Online Tasks},
 urldate = {2025-04-16},
 volume = {31},
 year = {2007}
}

@article{deFreitasNetto2020ConceptsReview,
    title = {{Concepts and forms of greenwashing: A systematic review}},
    year = {2020},
doi = {10.1186/s12302-020-0300-3},
    journal = {Environmental Sciences Europe},
    author = {de Freitas Netto, Sebastião Vieira and Sobral, Marcos Felipe Falcão and Ribeiro, Ana Regina Bezerra and Soares, Gleibson Robert da Luz},
    number = {1},
    pages = {19},
    volume = {32},
    publisher = {SpringerOpen}
}

@article{Moon2004ConsumerHypotheses,
    title = {{Consumer adoption of the internet as an information search and product purchase channel: some research hypotheses}},
    year = {2004},
    journal = {International Journal of Internet Marketing and Advertising},
    author = {Moon, Byeong-Joon},
    number = {1},
    month = {1},
    pages = {104--118},
    volume = {1},
    publisher = {Inderscience Publishers},
    url = {https://www.inderscienceonline.com/doi/abs/10.1504/IJIMA.2004.003692},
    doi = {10.1504/IJIMA.2004.003692},
    issn = {1477-5212}
}

@article{Davies2016ConsumerConsumption,
    title = {{Consumer motivations for mainstream “ethical” consumption}},
    year = {2016},
    journal = {European Journal of Marketing},
    author = {Davies, Iain Andrew and Gutsche, Sabrina},
    number = {7/8},
    month = {1},
    pages = {1326--1347},
    volume = {50},
    publisher = {Emerald Group Publishing Limited},
    url = {https://doi.org/10.1108/EJM-11-2015-0795},
    doi = {10.1108/EJM-11-2015-0795},
    issn = {0309-0566}
}

@article{Bloch1986ConsumerFramework,
 ISSN = {00935301, 15375277},
 URL = {http://www.jstor.org/stable/2489291},
 author = {Peter H. Bloch and Daniel L. Sherrell and Nancy M. Ridgway},
 journal = {Journal of Consumer Research},
 number = {1},
 pages = {119--126},
 publisher = {Oxford University Press},
 title = {Consumer Search: An Extended Framework},
 urldate = {2025-04-16},
 volume = {13},
 year = {1986}
}

@article{Annunziata2011ConsumersProducts,
    title = {{Consumers' attitudes toward labelling of ethical products: The case of organic and fair trade products}},
    year = {2011},
    journal = {Journal of Food Products Marketing},
    author = {Annunziata, Azzurra and Ianuario, Sara and Pascale, Paola},
    number = {5},
    month = {10},
    pages = {518--535},
    volume = {17},
    doi = {10.1080/10454446.2011.618790},
    issn = {10454446}
}

@book{Engel1990CustomerBehavior,
title = {Consumer behavior},
    year = {1990},
author = {Engel, James F and Blackwell, Roger D and Miniard, Paul W},
address = {Hinsdale, IL},
edition = {Sixth},
isbn = {0030018927},
language = {eng},
publisher = {Dryden Press},
}

@article{Kardes2006DebiasingNeglect,
    title = {{Debiasing omission neglect}},
    year = {2006},
    journal = {Journal of Business Research},
    author = {Kardes, Frank R and Posavac, Steven S and Silvera, David and Cronley, Maria L and Sanbonmatsu, David M and Schertzer, Susan and Miller, Felicia and Herr, Paul M and Chandrashekaran, Murali},
    number = {6},
    pages = {786--792},
    volume = {59},
    url = {https://www.sciencedirect.com/science/article/pii/S0148296306000324},
    doi = {https://doi.org/10.1016/j.jbusres.2006.01.016},
    issn = {0148-2963}
}

@inproceedings{Hearst2006DesignInterfaces,
    title = {{Design recommendations for hierarchical faceted search interfaces}},
    year = {2006},
    booktitle = {Proc. SIGIR 2006, Workshop on Faceted Search},
    author = {Hearst, M A},
    month = {8},
    pages = {26--30}
}

@article{Jansen2008DeterminingQueries,
title = {Determining the informational, navigational, and transactional intent of Web queries},
journal = {Information Processing \& Management},
volume = {44},
number = {3},
pages = {1251-1266},
year = {2008},
issn = {0306-4573},
doi = {https://doi.org/10.1016/j.ipm.2007.07.015},
url = {https://www.sciencedirect.com/science/article/pii/S030645730700163X},
author = {Bernard J. Jansen and Danielle L. Booth and Amanda Spink}
}

@inproceedings{Kallumadi2023ECom23:ECommerce,
author = {Kallumadi, Surya and Kim, Yubin and King, Tracy Holloway and Malmasi, Shervin and de Rijke, Maarten and Tagliabue, Jacopo},
title = {eCom'23: The SIGIR 2023 Workshop on eCommerce},
year = {2023},
isbn = {9781450394086},
publisher = {Association for Computing Machinery},
address = {New York, NY, USA},
url = {https://doi.org/10.1145/3539618.3591927},
doi = {10.1145/3539618.3591927},
booktitle = {Proceedings of the 46th International ACM SIGIR Conference on Research and Development in Information Retrieval},
pages = {3476–3478},
numpages = {3},
location = {Taipei, Taiwan},
series = {SIGIR '23}
}

@article{Pfeiffer2014EffectsEvaluation,
    title = {{Effects of Construal Level on Omission Detection and Multiattribute Evaluation}},
    year = {2014},
    journal = {Psychology {\&} Marketing},
    author = {Pfeiffer, Bruce E and Deval, Hélène and Kardes, Frank R and Ewing, Douglas R and Han, Xiaoqi and Cronley, Maria L},
    number = {11},
    month = {11},
    pages = {992--1007},
    volume = {31},
    publisher = {John Wiley {\&} Sons, Ltd},
    url = {https://doi.org/10.1002/mar.20748},
    doi = {https://doi.org/10.1002/mar.20748},
    issn = {0742-6046}
}

@article{Keller1987EffectsEffectiveness,
    title = {{Effects of Quality and Quantity of Information on Decision Effectiveness}},
    year = {1987},
    journal = {Journal of Consumer Research},
    author = {Keller, Kevin Lane and Staelin, Richard},
    number = {2},
    month = {9},
    pages = {200--213},
    volume = {14},
    url = {https://doi.org/10.1086/209106},
    doi = {10.1086/209106},
    issn = {0093-5301}
}

@article{Wiederhold2018EthicalIndustry,
    title = {{Ethical consumer behaviour in Germany: The attitude-behaviour gap in the green apparel industry}},
    year = {2018},
    journal = {International Journal of Consumer Studies},
    author = {Wiederhold, Marie and Martinez, Luis F.},
    number = {4},
    month = {7},
    pages = {419--429},
    volume = {42},
    publisher = {Blackwell Publishing Ltd},
    doi = {10.1111/ijcs.12435},
    issn = {14706431}
}

@article{Uusitalo2004EthicalFinland,
    title = {{Ethical consumerism: A view from Finland}},
    year = {2004},
    journal = {International Journal of Consumer Studies},
    author = {Uusitalo, Outi and Oksanen, Reetta},
    number = {3},
    pages = {214--221},
    volume = {28},
    doi = {10.1111/j.1470-6431.2003.00339.x},
    issn = {14706431}
}

@article{Ghose2014ExaminingRevenue,
    title = {{Examining the Impact of Ranking on Consumer Behavior and Search Engine Revenue}},
    year = {2014},
    journal = {Management Science},
    author = {Ghose, Anindya and Ipeirotis, Panagiotis G and Li, Beibei},
    number = {7},
    month = {2},
    pages = {1632--1654},
    volume = {60},
    publisher = {INFORMS},
    url = {https://doi.org/10.1287/mnsc.2013.1828},
    doi = {10.1287/mnsc.2013.1828},
    issn = {0025-1909}
}

@article{Djafarova2022ExploringBehaviour,
    title = {{Exploring ethical consumption of generation Z: theory of planned behaviour}},
    year = {2022},
    journal = {Young Consumers},
    author = {Djafarova, Elmira and Foots, Sophie},
    number = {3},
    month = {7},
    pages = {413--431},
    volume = {23},
    publisher = {Emerald Group Holdings Ltd.},
    doi = {10.1108/YC-10-2021-1405},
    issn = {17587212}
}

@article{Shaw2006FashionChoice,
    title = {{Fashion victim: The impact of fair trade concerns on clothing choice}},
    year = {2006},
    journal = {Journal of Strategic Marketing},
    author = {Shaw, Deirdre and Hogg, Gillian and Wilson, Elaine and Shiu, Edward and Hassan, Louise},
    number = {4},
    pages = {427--440},
    volume = {14},
    doi = {10.1080/09652540600956426},
    issn = {14664488}
}

@article{Brynjolfsson1999FrictionlessRetailers,
author = {Brynjolfsson, Erik and Smith, Michael D.},
title = {Frictionless Commerce? A Comparison of Internet and Conventional Retailers},
year = {2000},
url = {https://www.jstor.org/stable/2661602},
issue_date = {April 2000},
publisher = {INFORMS},
address = {Linthicum, MD, USA},
volume = {46},
number = {4},
issn = {0025-1909},
journal = {Manage. Sci.},
month = apr,
pages = {563–585},
numpages = {23}
}

@incollection{Ajzen1985FromBehavior,
    title = {{From Intentions to Actions: A Theory of Planned Behavior}},
    year = {1985},
    booktitle = {Action Control: From Cognition to Behavior},
    author = {Ajzen, Icek},
    editor = {Kuhl, Julius and Beckmann, Jürgen},
    pages = {11--39},
    doi = {10.1007/978-3-642-69746-3_2}
}

@inproceedings{Liu2020IdentifyingTasks,
    title = {{Identifying and predicting the states of complex search tasks}},
    year = {2020},
    booktitle = {Proceedings of the 2020 Conference on Human Information Interaction and Retrieval},
series = {{CHIIR} `20},
    author = {Liu, Jiqun and Sarkar, Shawon and Shah, Chirag},
    month = {3},
    pages = {193--202},
    _publisher = {Association for Computing Machinery, Inc},
    isbn = {9781450368926},
    doi = {10.1145/3343413.3377976},
    keywords = {Complex search task, Interactive ir, Task state}
}

@article{Ozcaglar-Toulouse2006InFrance,
    title = {{In search of fair trade: Ethical consumer decision making in France}},
    year = {2006},
    journal = {International Journal of Consumer Studies},
    author = {Ozcaglar-Toulouse, Nil and Shiu, Edward and Shaw, Deirdre},
    number = {5},
    pages = {502--514},
    volume = {30},
    doi = {10.1111/j.1470-6431.2006.00532.x},
    issn = {14706431}
}

@article{Schleenbecker2015InformationCoffee,
    title = {{Information needs for a purchase of fairtrade coffee}},
    year = {2015},
    journal = {Sustainability (Switzerland)},
    author = {Schleenbecker, Rosa and Hamm, Ulrich},
    number = {5},
    pages = {5944--5962},
    volume = {7},
    publisher = {MDPI},
    doi = {10.3390/su7055944},
    issn = {20711050}
}

@article{Zander2012InformationFood,
    title = {{Information search behaviour and its determinants: The case of ethical attributes of organic food}},
    year = {2012},
    journal = {International Journal of Consumer Studies},
    author = {Zander, Katrin and Hamm, Ulrich},
    number = {3},
    month = {5},
    pages = {307--316},
    volume = {36},
    doi = {10.1111/j.1470-6431.2011.00998.x},
    issn = {14706423}
}

@article{WardInternetBranding,
title = {Internet shopping, consumer search and product branding},
author = {Ward, Michael R. and Lee, Michael J.},
pages = {6-20},
_url = {https://www.emerald.com/insight/content/doi/10.1108/10610420010316302/full/html},
year = {2000},
month = {feb},
_urldate = {2025-02-11},
journal = {Journal of Product \& Brand Management},
volume = {9},
number = {1},
issn = {1061-0421},
doi = {10.1108/10610420010316302}
}

@inproceedings{Smith2019Knowledge-ContextActions,
    title = {{Knowledge-Context in Search Systems: Toward Information-Literate Actions}},
    year = {2019},
    booktitle = {Proceedings of the 2019 Conference on Human Information Interaction and Retrieval - CHIIR '19},
    author = {Smith, Catherine L and Rieh, Soo Young},
    pages = {55--62},
    _publisher = {ACM Press},
    _address = {New York, New York, USA}
}

@article{Papaoikonomou2018LookingApproach,
    title = {{Looking for info? Understanding ethical consumer information management using a diary approach}},
    year = {2018},
    journal = {Management Decision},
    author = {Papaoikonomou, Eleni and Valor, Carmen and Ginieis, Matias},
    number = {3},
    month = {3},
    pages = {645--662},
    volume = {56},
    publisher = {Emerald Group Holdings Ltd.},
    doi = {10.1108/MD-11-2016-0761},
    issn = {00251747}
}

@article{CarringtonLostGap,
title = {Lost in translation: Exploring the ethical consumer intention–behavior gap},
journal = {Journal of Business Research},
volume = {67},
number = {1},
pages = {2759-2767},
year = {2014},
issn = {0148-2963},
doi = {https://doi.org/10.1016/j.jbusres.2012.09.022},
url = {https://www.sciencedirect.com/science/article/pii/S0148296312002597},
author = {Michal J. Carrington and Benjamin A. Neville and Gregory J. Whitwell}
}

@article{Sheth2011MindfulSustainability,
  author    = {Jagdish N. Sheth and Nirmal K. Sethia and Shanthi Srinivas},
  title     = {Mindful consumption: a customer-centric approach to sustainability},
  journal   = {Journal of the Academy of Marketing Science},
  volume    = {39},
  number    = {1},
  pages     = {21--39},
  year      = {2011},
  month     = {February},
  doi       = {10.1007/s11747-010-0216-3},
  _url       = {https://doi.org/10.1007/s11747-010-0216-3},
  issn      = {1552-7824}
}

@inproceedings{Choi2021OrgBox:Tasks,
author = {Choi, Bogeum and Arguello, Jaime and Capra, Robert and Ward, Austin R.},
title = {OrgBox: A Knowledge Representation Tool to Support Complex Search Tasks},
year = {2021},
_isbn = {9781450380553},
_publisher = {Association for Computing Machinery},
_address = {New York, NY, USA},
_url = {https://doi.org/10.1145/3406522.3446029},
doi = {10.1145/3406522.3446029},
abstract = {Current search systems are effective in helping users complete simple search tasks (e.g., fact-finding). However, they provide less support for users completing complex search tasks. Complex search tasks involve a diverse set of cognitive and metacognitive activities, such as goal-setting, organizing information, drawing inferences, monitoring progress, and updating mental models. We report on a lab study (N=32) that investigated the uses and influences of a novel knowledge representation tool called the "OrgBox'', developed to support searchers with complex tasks. The OrgBox was integrated into a custom-built search system and allowed participants to save information by drag-and-dropping textual passages into the tool, organize passages into "boxes'', and make notes on passages and boxes. The OrgBox tool was compared to a baseline tool (the "Bookmark'') that allowed participants to save passages, but not organize them nor make notes. We investigate four research questions. In RQ1, we investigate the effects of the knowledge representation tool on participants' post-task perceptions. In RQ2-RQ4, we investigate: (RQ2) how participants used different features of each tool; (RQ3) the perceived benefits and challenges of each tool; and (RQ4) the influences of each tool on the approaches taken by participants to complete the task. To address RQ2-RQ4, we conducted a qualitative analysis of participants' responses during an exit interview. We discuss implications from our results for designing tools to support users with complex search tasks.},
booktitle = {Proceedings of the 2021 Conference on Human Information Interaction and Retrieval},
pages = {219–228},
numpages = {10},
_keywords = {search assistance tools, knowledge representation, complex tasks},
_location = {Canberra ACT, Australia},
series = {CHIIR '21}
}

@article{RowleyProductPropositions,
title = {Product search in e‐shopping: a review and research propositions},
author = {Rowley, Jennifer},
pages = {20-35},
_url = {https://www.emerald.com/insight/content/doi/10.1108/07363760010309528/full/html},
year = {2000},
month = {feb},
day = {1},
urldate = {2025-02-11},
journal = {Journal of Consumer Marketing},
volume = {17},
number = {1},
issn = {0736-3761},
doi = {10.1108/07363760010309528},
abstract = {The first stage in the consumer buying process is generally recognised to be that of the information search. This stage is recognised to be an important phase during which promotional messages should reach the intending consumer. Like many other stages of the buying process information seeking becomes more structured and constrained in the e‐shopping environment. In particular, the ability to collect product information and make comparisons between the different product offerings from different providers, possibly across national and currency boundaries, is often viewed as one of the main competitive challenges of e‐shopping. This article first visits models of the consumer buying process. It then explores the two approaches to information seeking, browsing and directed searching, and then proceeds to identify the tools that support these approaches. The complexity of variety in these tools is explored, in order to set the scene for understanding the complexity of the options with which the shopper is confronted.}
}

@article{Ke2016SearchProducts,
    title = {{Search for information on multiple products}},
    year = {2016},
    journal = {Management Science},
    author = {Ke, T. Tony and Shen, Zuo Jun Max and Villas-Boas, J. Miguel},
    number = {12},
    month = {12},
    pages = {3576--3603},
    volume = {62},
    publisher = {INFORMS Inst.for Operations Res.and the Management Sciences},
    doi = {10.1287/mnsc.2015.2316},
    issn = {15265501}
}

@incollection{dunning2011dunning,
title = {Chapter five - The Dunning–Kruger Effect: On Being Ignorant of One's Own Ignorance},
editor = {James M. Olson and Mark P. Zanna},
booktitle = {Advances in Experimental Social Psychology},
_publisher = {Academic Press},
_address = {Cambridge, MA, USA},
volume = {44},
pages = {247-296},
year = {2011},
_issn = {0065-2601},
doi = {https://doi.org/10.1016/B978-0-12-385522-0.00005-6},
_url = {https://www.sciencedirect.com/science/article/pii/B9780123855220000056},
author = {David Dunning},
keywords = {Ignorance, Self-evaluation, Known knowns, Known unknowns, Unknown unknowns, Overconfidence, Incompetence, Self-enhancement bias, Self-assessment},
abstract = {In this chapter, I provide argument and evidence that the scope of people's ignorance is often invisible to them. This meta-ignorance (or ignorance of ignorance) arises because lack of expertise and knowledge often hides in the realm of the “unknown unknowns” or is disguised by erroneous beliefs and background knowledge that only appear to be sufficient to conclude a right answer. As empirical evidence of meta-ignorance, I describe the Dunning–Kruger effect, in which poor performers in many social and intellectual domains seem largely unaware of just how deficient their expertise is. Their deficits leave them with a double burden—not only does their incomplete and misguided knowledge lead them to make mistakes but those exact same deficits also prevent them from recognizing when they are making mistakes and other people choosing more wisely. I discuss theoretical controversies over the interpretation of this effect and describe how the self-evaluation errors of poor and top performers differ. I also address a vexing question: If self-perceptions of competence so often vary from the truth, what cues are people using to determine whether their conclusions are sound or faulty?}
}

@article{belkin1982,
title = {{ASK} for information retrieval: Part I. Background and theory},
author = {Belkin, N.J. and Oddy, R.N. and Brooks, H.M.},
pages = {61-71},
url = {http://www.emeraldinsight.com/doi/10.1108/eb026722},
year = {1982},
month = {feb},
urldate = {2018-02-27},
journal = {Journal of Documentation},
volume = {38},
number = {2},
issn = {0022-0418},
doi = {10.1108/eb026722}
}

@article{Hasanzade2018SelectingShopping,
    title = {{Selecting decision-relevant ethical product attributes for grocery shopping}},
    year = {2018},
    journal = {Management Decision},
    author = {Hasanzade, Vüsal and Osburg, Victoria Sophie and Toporowski, Waldemar},
    number = {3},
    month = {3},
    pages = {591--609},
    volume = {56},
    publisher = {Emerald Group Holdings Ltd.},
    doi = {10.1108/MD-12-2016-0946},
    issn = {00251747}
}

@inproceedings{Awadallah2014SupportingTasks,
    title = {{Supporting complex search tasks}},
    year = {2014},
    booktitle = {CIKM 2014 - Proceedings of the 2014 ACM International Conference on Information and Knowledge Management},
    author = {Awadallah, Ahmed Hassan and White, Ryen W. and Pantel, Patrick and Dumais, Susan T. and Wang, Yi Min},
    month = {11},
    pages = {829--838},
    doi = {10.1145/2661829.2661912}
}

@inproceedings{Crescenzi2021SupportingOrgBox,
author = {Crescenzi, Anita and Ward, Austin R. and Li, Yuan and Capra, Rob},
title = {Supporting Metacognition during Exploratory Search with the OrgBox},
year = {2021},
_isbn = {9781450380379},
_publisher = {Association for Computing Machinery},
_address = {New York, NY, USA},
_url = {https://doi.org/10.1145/3404835.3462955},
doi = {10.1145/3404835.3462955},
abstract = {Current search systems provide effective support to users engaged in fact-finding and look-up oriented tasks. However, they provide relatively little support for users engaged in exploratory search tasks that involve cognitive and metacognitive activities such as learning, synthesis, planning, and reflection. We conducted a within-subject user study (N=24) that investigated the effects of a novel knowledge organization tool called the OrgBox, designed to assist users with organizing and synthesizing information, and metacognitive activities. The OrgBox included features to allow users to drag-drop information they found through search into "boxes" that could be created, labelled, and re-arranged. Study participants completed two exploratory search tasks, one with the OrgBox, and one with the OrgDoc, a baseline tool that included features of a rich-text editor (e.g., formatting, bullets) for taking notes.In this paper, we present results from our study comparing the OrgBox and OrgDoc tools. Specifically, we investigate if there were differences in participants' (1) search interactions, (2) saving and organizing behaviors (e.g., amount of information, structure of notes), (3) perceptions of the tasks, tool usability, and quality of their task outputs, and (4) perceptions of how the tools provided support for cognitive and metacognitive activities involved in the task. Our results show that when using the OrgBox tool, participants created more grouping sections in their notes and saved more text. In terms of metacognitive support, participants perceived the OrgBox tool to provide significantly higher levels of support for three types of metacognitive activity (monitoring/tracking, evaluation, and planning) without changing their perceptions of the task difficulty.},
booktitle = {Proceedings of the 44th International ACM SIGIR Conference on Research and Development in Information Retrieval},
pages = {1197–1207},
numpages = {11},
keywords = {exploratory search, knowledge organization tool, metacognition, orgbox},
_location = {Virtual Event, Canada},
series = {SIGIR '21}
}

@article{Young2010SustainableProducts,
    title = {{Sustainable consumption: Green consumer behaviour when purchasing products}},
    year = {2010},
    journal = {Sustainable Development},
    author = {Young, William and Hwang, Kumju and McDonald, Seonaidh and Oates, Caroline J.},
    number = {1},
    month = {1},
    pages = {20--31},
    volume = {18},
    doi = {10.1002/sd.394},
    issn = {09680802}
}

@article{brandtzaeg2017,
title = {Trust and distrust in online fact-checking services},
author = {Brandtzaeg, Petter Bae and F{\o}lstad, Asbj{\o}rn},
pages = {65-71},
url = {http://dl.acm.org/citation.cfm?doid=3134526.3122803},
year = {2017},
month = {aug},
day = {23},
urldate = {2023-10-31},
journal = {Communications of the {ACM}},
volume = {60},
number = {9},
issn = {00010782},
doi = {10.1145/3122803}
}

@article{sanbonmatsu2003,
title = {Overestimating the importance of the given information in multiattribute consumer judgment},
author = {Sanbonmatsu, David M. and Kardes, Frank R. and Houghton, David C. and Ho, Edward A. and Posavac, Steven S.},
pages = {289-300},
url = {http://doi.wiley.com/10.1207/{S15327663JCP1303\_10}},
year = {2003},
month = {jan},
urldate = {2021-02-11},
journal = {Journal of Consumer Psychology},
volume = {13},
number = {3},
issn = {10577408},
doi = {10.1207/S15327663JCP1303\_10}
}

@book{Jones2022TheDifference,
    title = {{The Better World Shopping Guide:: Every Dollar Makes a Difference}},
    year = {2022},
    author = {Jones, Ellis},
    publisher = {New Society Publishers}
}

@inproceedings{Choi2019TheComplexity,
author = {Choi, Bogeum and Capra, Robert and Arguello, Jaime},
title = {The Effects of Working Memory during Search Tasks of Varying Complexity},
year = {2019},
isbn = {9781450360258},
publisher = {Association for Computing Machinery},
address = {New York, NY, USA},
url = {https://doi.org/10.1145/3295750.3298948},
doi = {10.1145/3295750.3298948},
booktitle = {Proceedings of the 2019 Conference on Human Information Interaction and Retrieval},
pages = {261–265},
numpages = {5},
location = {Glasgow, Scotland UK},
series = {CHIIR '19}
}

@article{Liu2018TheIntention,
    title = {The influence of consumer mindset and corporate social responsility on purchase intention},
    year = {2018},
    journal = {Social Behavior and Personality: an international journal},
    author = {Liu, Xiaoping and Mao, Lijing and Deng, Wenxiang},
    number = {10},
    pages = {1647--1656},
doi = {10.2224/sbp.7025},
    volume = {46},
    publisher = {Scientific Journal Publishers}
}

@article{Casais2022ThePriorities,
    title = {{The Intention-Behavior gap in Ethical Consumption: Mediators, Moderators and Consumer Profiles Based on Ethical Priorities}},
    year = {2022},
    journal = {Journal of Macromarketing},
    author = {Casais, Beatriz and Faria, Joana},
    number = {1},
    month = {3},
    pages = {100--113},
    volume = {42},
    publisher = {SAGE Publications Inc.},
    doi = {10.1177/02761467211054836},
    issn = {15526534}
}

@article{Akerlof1970TheMechanism,
title = {The Market for "Lemons": Quality Uncertainty and the Market Mechanism},
author = {Akerlof, George A.},
pages = {488--500},
url = {http://qje.oxfordjournals.org/lookup/doi/10.2307/1879431},
year = {1970},
month = {aug},
urldate = {2025-04-16},
journal = {The quarterly journal of economics},
volume = {84},
number = {3},
issn = {00335533},
doi = {10.2307/1879431}
}

@article{Ellen1991TheBehaviors,
    title = {{The Role of Perceived Consumer Effectiveness in Motivating Environmentally Conscious Behaviors}},
    year = {1991},
    journal = {Journal of Public Policy {\&} Marketing},
    author = {Ellen, Pam and Wiener, Joshua and Cobb-Walgren, Cathy},
    month = {9},
    pages = {102--117},
    volume = {10},
    doi = {10.1177/074391569101000206}
}

@article{Wildemuth2014UntanglingStudies,
    title = {{Untangling search task complexity and difficulty in the context of interactive information retrieval studies}},
    year = {2014},
    journal = {Journal of Documentation},
    author = {Wildemuth, Barbara and Freund, Luanne and G. Toms, Elaine},
    editor = {Peter Willett, Professor},
    number = {6},
    month = {1},
    pages = {1118--1140},
    volume = {70},
    publisher = {Emerald Group Publishing Limited},
    url = {https://doi.org/10.1108/JD-03-2014-0056},
    doi = {10.1108/JD-03-2014-0056},
    issn = {0022-0418}
}

\end{document}